\documentclass[twocolumn,showpacs,preprintnumbers,amsmath,amssymb,floatfix,prd]{revtex4}
\usepackage{epsfig}
\usepackage{dcolumn}
\begin{document}
\title{Global Analysis of Nuclear Parton Distributions}
\author{Daniel de Florian}\email{deflo@df.uba.ar}
\affiliation{ Departamento de F\'{\i}sica and IFIBA,  Facultad de Ciencias Exactas y Naturales, Universidad de Buenos Aires, Ciudad Universitaria, Pabell\'on\ 1 (1428) Buenos Aires, Argentina}
\affiliation{  Institut f\"ur Theoretische Physik, Universit\"at Z\"urich, CH-8057 Z\"urich, Switzerland}
\author{Rodolfo Sassot}\email{sassot@df.uba.ar}
\affiliation{ Departamento de F\'{\i}sica and IFIBA,  Facultad de Ciencias Exactas y Naturales, Universidad de Buenos Aires, Ciudad Universitaria, Pabell\'on\ 1 (1428) Buenos Aires, Argentina}
\author{Marco Stratmann}\email{marco@bnl.gov}
\affiliation{Physics Department, Brookhaven National Laboratory, Upton, NY~11973, USA}
\author{Pia Zurita}\email{pia@df.uba.ar}
\affiliation{ Departamento de F\'{\i}sica and IFIBA,  Facultad de Ciencias Exactas y Naturales, Universidad de Buenos Aires, Ciudad Universitaria, Pabell\'on\ 1 (1428) Buenos Aires, Argentina}

\begin{abstract}
We present a new global QCD analysis of nuclear parton distribution functions
and their uncertainties.
In addition to the most commonly analyzed data sets
for the deep-inelastic scattering of charged leptons off nuclei and 
Drell-Yan di-lepton production, we include also measurements for neutrino-nucleus
scattering and inclusive pion production in deuteron-gold collisions.
The analysis is performed at next-to-leading order accuracy in perturbative QCD in
a general mass variable flavor number scheme, adopting a current set of free nucleon parton 
distribution functions, defined accordingly, as reference. 
The emerging picture is one of consistency, where universal 
nuclear modification factors for each parton flavor reproduce the main features 
of all data without any significant tension among the different sets.
We use the Hessian method to estimate the uncertainties of the obtained nuclear modification
factors and examine critically their range of validity in view of the sparse kinematic
coverage of the present data.
We briefly present several applications of our nuclear parton densities in 
hard nuclear reactions at BNL-RHIC, CERN-LHC, and a future electron-ion collider.
\end{abstract}


\maketitle

\section{Introduction and Motivation}
%
In spite of the remarkable phenomenological success of Quantum Chromodynamics
(QCD) as the theory of strong interactions, a detailed understanding of the role 
of quark and gluon degrees of freedom in nuclear matter is still lacking and 
subject to ongoing experimental and theoretical efforts. 
In this context, the rather unexpected discovery, 
almost three decades ago, that quarks and gluons in bound nucleons exhibit 
non-trivial momentum distributions, noticeably different from those measured in free or 
loosely bound nucleons \cite{Aubert:1983xm}, has triggered a long-standing quest 
for more and more precise determinations of nuclear parton distribution functions (nPDFs). 
These endeavors have led to increasingly 
accurate and comprehensive measurements of cross sections involving different
hard scattering processes and nuclear targets \cite{Arneodo:1992wf},
a better theoretical insight into the underlying physics, 
and a more refined framework for analyses of nPDFs \cite{ref:nds,ref:hirai,ref:eps09}.

On the one hand, a reliable extraction of nPDFs from data is required for a deeper understanding of 
the mechanisms associated with nuclear binding from a QCD improved parton model 
perspective, including a verification of various proposed nuclear modifications
whose phenomenological details vary from model to model \cite{ref:models},
leading to a wide spread of expectations.
On the other hand, nPDFs are a vital input for the theoretical interpretation
and analyses of a large variety of ongoing and future high energy nuclear 
physics experiments, such as, for instance, heavy ion collisions at BNL-RHIC, 
proton-nucleus collisions to be performed at CERN-LHC \cite{Accardi:2004be,arXiv:1105.3919},  
or deep-inelastic neutrino-nucleus interactions in long baseline neutrino 
experiments \cite{Paschos:2001np}. 
Another important physics objective related to nPDFs is to explore and 
quantify the effects of multiple rescatterings and recombinations of 
small momentum fraction gluons, leading to deviations \cite{ref:nonlinear} from the linear scale evolution 
usually assumed for nPDFs.
The transition to the saturation regime is often characterized by the saturation
scale $Q_s$ which depends on both the relevant momentum fraction $x$ and the atomic number $A$.
Important quantitative benchmark tests of saturation phenomena can be performed at
a future electron-heavy ion collider \cite{ref:lhec,ref:eic}.
As a result, the kinematic range and the
accuracy at which nPDFs are known will continue to be a topical issue 
in many areas of high energy nuclear physics.

In the last few years, significant progress has been made in obtaining nPDFs from data. 
In addition to the theoretical improvements nowadays routinely used in modern 
extractions of free proton PDFs,
such as the consistent implementation of QCD corrections beyond the leading 
order \cite{ref:nds} and uncertainty estimates \cite{ref:hirai,ref:eps09}, 
the most recent determinations of nPDFs have also extended the
types of data sets taken into account, moving towards truly 
global QCD analyses of nuclear effects \cite{ref:eps09,ref:schienbein,ref:paukkunen,Kovarik:2010uv}. 
The addition of novel hard probes to the fit does not only lead to better constrained sets of nPDFs
and allows one to study the nuclear modification to the different parton species individually,
but also tests the assumed process independence of nuclear effects.
Verifying the range of applicability of standard factorization theorems and the universality of nPDFs
for hard probes in nuclear collisions is of outmost importance as formally power suppressed 
``higher twist'' contributions can be much enhanced due to the larger density of gluons in heavy nuclei.

The deep-inelastic scattering (DIS) of charged leptons off nuclear targets
not only initiated all studies of nPDFs but still
provides the best constraints on nuclear modifications for quark distributions.
Current data comprise a wide selection of nuclei from helium to lead, are presented
as ratios of structure functions for two different nuclei, and roughly span the range
$0.01\lesssim x \lesssim 1$ of momentum fractions. 
Although a separation between quarks and antiquarks is not possible based on these data alone,
DIS at medium-to-large $x$ mainly probes valence quarks,
while data at lower momentum fractions $x\simeq 0.01$ are sensitive to 
nuclear modifications of sea quarks.
Upon combination with available data on Drell-Yan (DY) di-lepton production off nuclear targets,
a better discrimination between valence and sea quarks can be achieved,
mainly hampered, however, by large experimental uncertainties and the 
limited kinematic coverage.

DIS and DY data only loosely constrain the nuclear modifications to the gluon density
because they cover a too limited range in the hard energy scale $Q$ such that evolution
effects are small and at best comparable to the current experimental precision.
To remedy this situation and to further constrain the nuclear gluon density,
data from BNL-RHIC for inclusive pion production in deuteron-gold 
$(dAu)$ collisions have been included in the analysis of nPDFs performed in Ref.~\cite{ref:eps09}. 
Gluon initiated processes are known to be dominant in
inclusive hadron or jet production at RHIC at not too large transverse momenta $p_T$ \cite{ref:hadron}, 
and analogous data for polarized proton-proton collisions indeed provide the best constraint
on the helicity dependent gluon density \cite{ref:dssv}.
Not surprisingly,
the data for $dAu$ collisions at mid rapidity used in the fit in  Ref.~\cite{ref:eps09}
have a significant impact on their determination of the gluon distribution in a gold nucleus.
The corresponding nuclear modification for gluons at medium to large $x$ turned out  
to be much more pronounced than in previous estimates \cite{ref:nds,ref:hirai}
and also much larger than those found for all the other partonic species.

Another promising avenue for significant improvements in extractions of nPDFs is
neutrino induced DIS off iron and lead targets available from NuTeV \cite{Tzanov:2005kr},
CDHSW \cite{Berge:1989hr}, and CHORUS \cite{ref:CHORUS}.
These data receive their importance from their discriminating power between 
nuclear modifications for quarks and antiquarks and have been included in a series
of analyses in Refs.~\cite{ref:schienbein,Kovarik:2010uv}.
Unexpectedly, the correction factors obtained from neutrino scattering data 
are found to differ significantly both in shape and in magnitude from those extracted with 
the more traditional charged lepton probes \cite{ref:schienbein,Kovarik:2010uv}.
At variance with these results, Ref.~\cite{ref:paukkunen} confronts the 
neutrino DIS cross sections with nPDFs obtained in \cite{ref:eps09}
without any refitting and finds no apparent disagreement between the nuclear
effects obtained from different hard probes.

The global QCD analysis of nPDFs presented here
incorporates in a comprehensive way all of the above mentioned improvements. 
The resulting set of nPDFs at next-to-leading order (NLO) accuracy
supersedes previous work presented in \cite{ref:nds}.
The fitting procedure is efficiently performed in Mellin moment space based
on techniques presented and used in \cite{ref:mellin,ref:dssv}.
We adopt a contemporary set of free nucleon PDFs \cite{Martin:2009iq}
defined in leading-twist collinear factorization in the $\overline{\mathrm{MS}}$
scheme as our reference distribution to quantify modifications of PDFs in nuclei.
The same general mass variable flavor number scheme
(GM-VFNS) as in \cite{Martin:2009iq} is used in our analysis
to define charm and bottom quark contributions.

We utilize the Hessian method \cite{ref:hessian} 
to estimate the uncertainties of the nuclear modification
factors for quarks and gluons 
originating from the experimental errors on the fitted data points 
and examine critically their range of validity in view of the sparse kinematic
coverage of the present data.
The resulting eigenvector sets of nPDFs enable one to propagate uncertainties
to any desired observable. 
We also highlight interesting complications due the possibility of having 
negative parton densities beyond the leading order approximation at small values
of $x$ in the vicinity of typical initial scales of $1\,\mathrm{GeV}$ for PDF
evolution without spoiling the positivity of measured physical observables. 
Such a scenario is realized for our reference gluon density at NLO (and beyond)
in a free nucleon \cite{Martin:2009iq} and propagates also to the gluon distribution in nuclei
obtained in this analysis.

In addition to the neutral pion production data from 
PHENIX \cite{ref:phenixpi0} used in \cite{ref:eps09},
we include also the charged \cite{ref:starpipm}
and the recently published neutral pion \cite{ref:starpi0}
data from the STAR experiment in our fit.
Instead of adopting only vacuum parton-to-pion fragmentation functions (FFs),
such as the ones given in Ref.~\cite{ref:dsspion}, in our calculations,
we account for possible medium modifications in the formation of the pions
by utilizing also a set of nuclear FFs (nFFs) \cite{ref:nff} which reproduces the
large hadron attenuation observed in DIS multiplicities by the HERMES collaboration
\cite{Airapetian:2009jy}.
Whenever possible, we compare measured minimum bias cross sections in $dAu$ collisions
with our computations at NLO accuracy, rather than utilizing 
nuclear modification factors $R_{dAu}^{\pi}$
whose relation to cross sections introduces an additional model dependence.
Regarding the use of neutrino data, we include the charged current DIS structure 
functions $F_2^{\nu A}$ and $F_3^{\nu A}$ from NuTeV, CDHSW, and CHORUS for iron and 
lead targets \cite{Tzanov:2005kr,Berge:1989hr,ref:CHORUS} in our analysis. 
Mass effects for heavy quarks are consistently taken into account using the recently obtained
expressions of the NLO coefficients \cite{ref:nlocc} 
in Mellin moment space \cite{Blumlein:2011zu}. 
They are of particular relevance for a proper treatment of the strangeness contribution to
charged current DIS which produces a massive charm quark in the final state.

The main features of our new parametrization of nPDFs are worth emphasizing
already at this point. All current data can be described well within conventional
leading-twist collinear factorization at NLO accuracy by a universal set of nPDFs.
There are no indications yet for the onset of non-linear effects in the scale evolution
of nPDFs or a breakdown of factorization for hard probes involving one heavy nuclei. 
This is not too surprising given the limited kinematic coverage of the data, in particular,
with respect to the momentum fraction $x$.
We find neither the unusually large nuclear modifications of the
gluon distribution at medium to large $x$ obtained in the analysis of \cite{ref:eps09} 
nor any tension or discrepancy between
charged and neutral current DIS results reported in \cite{ref:schienbein,Kovarik:2010uv}.
These differences with previous analyses are perhaps a good measure of 
some, usually disregarded uncertainties inherent to global QCD fits 
such as the applied data selection criteria, the different flexibility of 
parameterizations of nuclear modifications, the way of propagating experimental
uncertainties, or the neglect of certain theoretical ambiguities. 
All these issues need to be inspected more closely in the future
but likely require more precise data and further advances in theory to be resolved.

The remainder of the paper is organized as follows: in the next Section 
we briefly review the general framework for a global QCD analysis of nPDFs,
establish our conventions, and describe the strategy of how to parametrize nuclear
modifications of PDFs in nuclei.
In Sec.~III we proceed with a detailed discussion and presentation of the results
of our analysis. We assess and critically examine the uncertainties of nPDFs in
Sec.~\ref{sec:npdf}. Some expectations for future hard probes of nPDFs such as prompt photon and 
DY di-lepton production at RHIC and the LHC or hadron yields in DIS
are presented in Sec.~\ref{sec:future}. We summarize the main results of our analysis in Sec.~V.

\section{Framework}
%
Throughout this analysis, we make the usual assumption \cite{ref:nds,ref:hirai,ref:eps09}
that theoretical expressions for measured cross sections $d\sigma^A$ 
involving a nucleus $A$ factorize into calculable
partonic hard scattering cross sections $d\hat{\sigma}$, 
identical to those used for processes involving free nucleons,
and appropriate combinations of non-perturbative collinear 
parton densities and fragmentation functions.
If applicable, the latter quantities are subject to nuclear modifications and
will be denoted as $f^A_i$ and $D^{A,h}_i$, respectively.
Here, $i$ labels the parton flavor, and $h$ represents the hadron species produced in
the fragmentation process. 
The scale dependence of $f^A_i$ and $D^{A,h}_i$
is dictated by the proper factorization of collinear mass singularities and,
hence, will be governed by the same evolution equations and kernels as for free nucleons or
fragmentation in the vacuum. As a result, the entire nuclear modification resides in the
initial conditions for $f^A_i$ and $D^{A,h}_i$ at some low scale $Q_0\simeq 1\;\mathrm{GeV}$
and needs to be parametrized from data.

Applying factorization, the cross sections for DIS, DY, and pion production
off nuclear beams or targets relevant for our global analysis 
schematically read
\begin{eqnarray}
\label{eq:disxsec}
d\sigma^A_{\mathrm{DIS}} &=& \sum_i f_i^A \otimes d\hat{\sigma}_{i\gamma^*\to X}\;, \\
\label{eq:dyxsec}
d\sigma^A_{\mathrm{DY}}  &=& \sum_{ij} f^p_i \otimes f^A_j \otimes d\hat{\sigma}_{ij\rightarrow l\bar{l}X}\;,\\
\label{eq:daxsec}
d\sigma^A_{dA\to \pi X}  &=& \sum_{ijk} f^d_i \otimes f^A_j \otimes d\hat{\sigma}_{ij\rightarrow kX}
\otimes D^{A,\pi}_k \;,
\end{eqnarray}
respectively, where, for brevity, we have suppressed any 
dependence on kinematic variables, the strong coupling $\alpha_s$, renormalization, and factorization scales.
We note that the DIS cross section $d\sigma^A_{\mathrm{DIS}}$ in (\ref{eq:disxsec})
is usually expressed in terms of structure functions $F_{2,3,L}^{lA}$, where $l$ denotes
either a charged lepton or a neutrino, depending on the experimental setup. 
For the partonic hard scattering cross sections $d\hat{\sigma}$ in Eqs.~(\ref{eq:disxsec})-(\ref{eq:daxsec}),
the scale evolution of parton densities and fragmentation functions,
and the running of $\alpha_s$ we consistently use the available expressions at NLO accuracy 
in the $\overline{\mathrm{MS}}$ scheme throughout our analysis.
We refrain from performing a leading order extraction of nPDFs since 
this leads to a far inferior description of DY and pion production cross section data 
than in a NLO framework.

The factorization of all medium related effects into the initial conditions for 
the scale evolution of process independent nPDFs (and, if appropriate, nFFs)
is clearly an assumption and, apart from its phenomenological success in describing
current data, neither proven nor even expected to work in general.
It provides one, however, with a rigorous and testable calculational framework 
of great predictive power, order by order in perturbation theory. 
Global analyses of nPDFs can help to reveal its limitations by
looking for potential tensions with data.
Various mechanisms can ultimately lead to a breakdown of 
leading-twist factorization once a regime
of dense, saturated gluons is reached \cite{ref:nonlinear}.
As a consequence, nuclear PDFs are usually applied only to a 
large class of hard probes where
a nucleus collides with a lepton, a nucleon, or a very light nucleus like
the deuteron rather than to interactions of two heavy nuclei which create
high gluon densities. 
Another reason for neglecting heavy-ion collisions in nPDF analyses 
is the complete lack of control of the experimentally important  
impact parameter or centrality dependence of the probes in
a nPDF based framework.

The symbol $\otimes$ in Eqs.~(\ref{eq:disxsec})-(\ref{eq:daxsec})
denotes a convolution integral with respect to the relevant momentum fraction.
To avoid these time consuming integrations in a fit to a large body of data,
we use the Mellin technique as outlined in Refs.~\cite{ref:mellin,ref:dssv}
which allows one to treat lengthy NLO expressions numerically efficient but
without resorting to any approximations.
The idea is to represent all non perturbative quantities in Eqs.~(\ref{eq:disxsec})-(\ref{eq:daxsec})
by their representations as Mellin inverses, for instance,
\begin{equation}
f^A_i(x) = \frac{1}{2\pi i} \int_{{\cal{C}}_N} x^{-N} f^A_i(N) dN\;,
\end{equation}
where ${\cal{C}}_N$ is a suitable contour in the complex $N$ plane that has an imaginary part
ranging from $-\infty$ to $+\infty$ and that intersects the real axis to the right of the
rightmost pole of $f^A_i(N)$.
Next, after reshuffling integrations in (\ref{eq:disxsec})-(\ref{eq:daxsec}), one can compute all quantities,
except the desired $f^A_i(N)$ but including the time-consuming $d\hat{\sigma}$, 
prior to the actual fit and store them in multi-dimensional look-up tables in Mellin space
\cite{ref:mellin,ref:dssv}.
This technique has been successfully exploited in various other global fits of parton
densities and fragmentation functions \cite{ref:dsspion,ref:dssv,ref:nff,ref:dssproton}.

At variance with our previous analysis in Ref.~\cite{ref:nds}, where the initial nPDFs 
at scale $Q_0$ were related to some set of proton distributions through a convolution
\begin{equation}
\label{eq:npdfconv}
f^A_i(x,Q_0) = \int_{x_{N}}^A \frac{dy}{y} W_i^A(y,Q_0) f_i^p\left( \frac{x_N}{y},Q_0 \right)
\end{equation}
with appropriately fitted weights $W_i^A$, we will work this time within a 
more conventional approach \cite{ref:hirai,ref:eps09}
that defines the nPDFs for a bound proton in a nucleus $A$, $f^A_i$, 
with respect those for a free proton, $f^p_i$, 
through a multiplicative nuclear modification factor $R^A_i(x_N,Q_0)$ 
as
\begin{equation}
 f^A_i(x_N,Q_0) = R^A_i(x_N,Q_0)\, f_i^p (x_N,Q_0)\;.
\label{eq:npdfdef}
\end{equation}
$x_N$ resembles the usual DIS scaling variable for free nucleons, assuming that the
momentum of the nucleus $p_A$ is distributed evenly among its nucleons, i.e.,
$p_N=p_A/A$, and has, in principle, support in the interval $0<x_N<A$. This reflects the
fact that a parton in a nucleus may carry more than the average nucleon momentum $p_N$.
Since the $f_i^p$ are restricted to the range $0<x_N<1$, the nPDFs defined through Eq.~(\ref{eq:npdfdef}) 
are also constrained to $x_N<1$. Apart from being well suited to Mellin moment space,
the convolution approach in (\ref{eq:npdfconv})
has the advantage to allow for defining nPDFs also beyond $x_N=1$, see \cite{ref:nds}.
To facilitate comparisons to other analyses \cite{ref:hirai,ref:eps09} and to emphasize that
the results of our fit are not a consequence of adopting a different approach, we 
choose, however, the ansatz (\ref{eq:npdfdef}) to define our input distributions, which has 
the additional advantage of making 
the effects of nuclear modifications more transparent than in a convolution with
a weight function $W_i^A$. 

As the reference PDFs for the free proton, $f_i^p$, we select the latest NLO set from
the MSTW global QCD analysis \cite{Martin:2009iq} which is defined in
a general mass variable flavor number scheme to deal with heavy quark 
mass effects.
The nPDFs are then obtained by (\ref{eq:npdfdef}) at an initial scale of $Q_0=1\,\mathrm{GeV}$
by determining the nuclear modification factors $R^A_i$ from data. 
Their evolution to scales $Q>Q_0$ follows the prescriptions of the GM-VFNS as
specified in \cite{Martin:2009iq}. This includes the same choices for the initial value
and the running of the strong coupling and the masses and thresholds for the heavy 
charm and bottom quarks.
Notice that the medium modified heavy quark distributions are generated perturbatively from
the gluon and light quark flavors, so there is no need to introduce any additional
free parameters for them.

Our strategy to parametrize the $R^A_i(x_N,Q_0)$ in Eq.~(\ref{eq:npdfdef})
is as follows:
as in previous analyses \cite{ref:nds,ref:hirai,ref:eps09}, we assume isospin invariance and
neglect any nuclear effects for the deuteron. 
Both valence quark distributions are assigned
the same nuclear modification factor $R^A_v(x_N,Q^2_0)$, which we parametrize as
\begin{eqnarray}
\nonumber
R^A_v(x,Q^2_0)  &=&  \epsilon_1 \, x^{\alpha_v} (1-x)^{\beta_1}  \times  \\ 
         & &   (1 + \epsilon_2 (1-x)^{\beta_2}) (1 + a_v  (1-x)^{\beta_3})\,,
\label{eq:rval}
\end{eqnarray}
and where we have dropped the subscript $N$ in the parton momentum fraction variable
for simplicity.
The flexible functional form in (\ref{eq:rval}) can account for the typical $x$ dependent
pattern of nuclear corrections found in ratios of DIS structure functions for different nuclei such as
shadowing, anti-shadowing, EMC, and Fermi motion effects.

We also assume that the light sea quarks and antiquarks share the same correction factor 
$R^A_s(x,Q^2_0)$. No significant improvement in the quality of 
the fit to data is found by relaxing this assumptions and assigning different correction 
factors for each quark flavor. This is not too surprising given the limited kinematic coverage
and precision of the data. We choose another factor $R^A_g(x,Q^2_0)$ to parametrize 
medium effects for gluons.
Actually, it turns out that all current data are very well reproduced by using 
nuclear modification factors $R^A_i$ that are not completely independent.
An excellent description of the data is achieved by relating both $R_s^A$ and $R_g^A$ to
$R_v^A$ specified in Eq.~(\ref{eq:rval}), 
allowing only for a different normalization and modifications in the 
low-$x$ behavior. Hence we choose, without any loss in the quality of the fit, 
\begin{eqnarray}
 R^A_s(x,Q^2_0)  &=& R^A_v(x,Q^2_0) \,\frac{\epsilon_s}{\epsilon_1} \frac{1 + a_s x^{\alpha_s}}{a_s + 1}\;, \\
\label{eq:rsea}
 R^A_g(x,Q^2_0)  &=& R^A_v(x,Q^2_0) \,\frac{\epsilon_g}{\epsilon_1} \frac{1 + a_g x^{\alpha_g}}{a_g + 1} \;.
\label{eq:rglue}
\end{eqnarray}
At large $x$, where sea quark and gluon densities 
become very small compared to the dominant valence distributions, the fit cannot determine
extra parameters in their nuclear modifications individually. To a first approximation
it appears to be sensible to assume a common large $x$ behavior for all nPDFs.
It has to be kept in mind, however, that both the resulting EMC effect and the
Fermi motion for sea quarks and gluons at large $x$ are not a 
result of an experimental constraint but mere assumptions which have no impact on
the quality of the fit.
For the use of the Mellin technique outlined above, it is important to recall that
the $N$ moments of the $R_i^A$ defined in Eqs.~(\ref{eq:rval})-(\ref{eq:rglue}) can
be taken analytically, leading to appropriate combinations of Euler Beta functions.

To determine the number of actual fit parameters we note that the
coefficients $\epsilon_1$ and  $\epsilon_2$ in Eq.~(\ref{eq:rval}) 
are fixed by charge conservation, i.e.,
\begin{equation}
\int_0^1 dx\, f_{u_{v}}^A (x,Q^2)=2 \;\;\mathrm{and}\; \int_0^1 dx\, f_{d_{v}}^A (x,Q^2)=1 \;.
\label{eq:charge}
\end{equation}
This leaves one with nine free parameters per nucleus to reproduce all the features
of the DIS, DY, and $dAu$ data included in the fit,
if we further constrain $\epsilon_s$ and $\epsilon_g$ to be equal, which,
again, has no impact on the quality of the fit, and fix $\epsilon_s$ by
momentum conservation,
\begin{equation}
\int_0^1 dx\,\sum_i x\,f_{i}^A (x,Q^2)=1\;\;.
\label{eq:mom}
\end{equation}
The $A$ dependence of the remaining free parameters 
$\xi \in \{\alpha_v, \alpha_s,\alpha_g, \beta_1, \beta_2, \beta_3, a_v, a_s, a_g \}$ 
is parametrized in the usual way \cite{ref:nds} as
\begin{equation}
\label{eq:adep}
\xi = \gamma_{\xi} + \lambda_{\xi} A^{\delta_{\xi}}\;.
\end{equation}
The very mild $A$ dependence found for some of the $\xi$'s
allows us to further reduce the number of additional parameters in
(\ref{eq:adep}) by setting $\delta_{a_g}=\delta_{a_s}$ and 
$\delta_{\alpha_g}=\delta_{\alpha_s}$, leaving a total of
25 free fit parameters.

The optimum values of these parameters are extracted from data by performing a 
minimization of an effective $\chi^2$ function that quantifies the goodness
of the fit to data for a given set of parameters.
Given the still sizable experimental uncertainties of the data sensitive to nPDFs,
we choose the simplest $\chi^2$ function,
\begin{equation}
\label{eq:chi2}
\chi^2 \equiv \sum_i \omega_{i} \,
\frac{ (d\sigma^{\mathrm{exp}}_i-d\sigma^{\mathrm{th}}_i)^2}
{\Delta^2_i} 
\end{equation} 
where each experimental result
$d\sigma^{\mathrm{exp}}_i$ is compared to its corresponding theoretical estimate
$d\sigma^{\mathrm{th}}$, weighted with the uncertainties $\Delta_i$
for each data point. The latter are simply estimated by adding
statistical and systematic errors in quadrature.
The sum in (\ref{eq:chi2}) runs over all data points $i$ included in the fit,
and $\omega_i$ allows one to give artificial weights to different data sets. We refrain from
using this option and set $\omega_i=1$. In addition, there appears to be no need
in our fit for introducing relative normalization shifts among different sets of data.
We postpone a detailed discussion on how we estimate uncertainties of our nPDFs to
Sec.~\ref{sec:npdf}.

\section{Discussion of the Results}
%
In this Section we discuss in detail the results obtained from 
our NLO global QCD fit to nuclear scattering data.
We start with presenting the parameters of the best fit and
the resulting $\chi^2$ values for each set of data. In the
following Subsections we discuss the individual probes, DIS,
DY, neutrino DIS, and $dAu$ collisions, included in the fit
and show comparisons between data and theory. We finish by
presenting the obtained nuclear modification factors $R_i^A$
and assessing their uncertainties
in Subsec.~\ref{sec:npdf}.

\subsection{Determination of the optimum fit}

The data analyzed comprise the classic EMC \cite{EMC}, NMC \cite{NMC1,NMC2,NMC3}, 
and SLAC E139 \cite{SLAC1} results for
ratios of the DIS structure function $F_2^A(x,Q^2)$ for various heavy nuclei 
to those for deuterium, lithium, or carbon, see Tab.~\ref{tab:expchi2}.
We impose a cut $Q^2>1\,\mathrm{GeV}^2$ on the data to ensure that perturbative 
QCD is applicable, and we are in the deep-inelastic regime.
We also include the DY di-lepton production data taken in proton-nucleus collisions
from the E772 \cite{e772} and E866 \cite{e866} collaborations, presented as ratios of cross
sections for various heavy nuclei to those for deuterium and beryllium, respectively.
Data for single inclusive hadron production in deuteron-gold collisions 
from the PHENIX \cite{ref:phenixpi0} and STAR \cite{ref:starpipm,ref:starpi0}
experiments are are taken into account for pions at mid rapidity and 
$p_T> 2\,\mathrm{GeV}$ where NLO QCD provides a good description of corresponding
$pp$ data \cite{ref:dsspion}. 
Finally, results for neutrino DIS off iron and lead nuclei from the NuTeV \cite{Tzanov:2005kr},
CDHSW \cite{Berge:1989hr}, and CHORUS \cite{ref:CHORUS} collaboration are included, 
again after imposing a cut $Q^2>1\,\mathrm{GeV}^2$.
The total number of 1579 data points considered in our analysis exceeds those included in
our previous fit \cite{ref:nds} by almost a factor of four, which shows how timely this
re-analysis is.

\begin{table}[th!]
\caption{\label{tab:expchi2} Total and individual
$\chi^2$ values for the data sets included in the fit. }
\begin{ruledtabular}
\begin{tabular}{llccc} 
measurement              & collaboration    &   ref.  & \# points  &$\chi^2$ \\ \hline
$F_2^{He}/F_2^{D}$ & NMC       &  \cite{NMC1}  & 17  & 18.18    \\ 
                   & E139      &  \cite{SLAC1} & 18  & 2.71     \\ 
$F_2^{Li}/F_2^{D}$ & NMC       &  \cite{NMC1}  & 17  & 17.35    \\
$F_2^{Li}/F_2^{D}$ $Q^2$ dep. & NMC   &  \cite{NMC1}  & 179 & 197.36   \\
$F_2^{Be}/F_2^{D}$ & E139      &  \cite{SLAC1} & 17  & 44.17    \\ 
$F_2^{C}/F_2^{D}$  & NMC       &  \cite{NMC1}  & 17  & 27.85    \\ 
                   & E139      &  \cite{SLAC1} & 7   & 9.66     \\ 
                   & EMC       &  \cite{EMC}   & 9   & 6.41     \\
$F_2^{C}/F_2^{D}$ $Q^2$ dep.& NMC    &  \cite{NMC1}  & 191 & 201.63   \\                  
$F_2^{Al}/F_2^{D}$ & E139      &  \cite{SLAC1} & 17  & 13.22    \\ 
$F_2^{Ca}/F_2^{D}$ & NMC       &  \cite{NMC1}  & 16  & 18.60    \\ 
                   & E139      &  \cite{SLAC1} & 7   & 12.13    \\
$F_2^{Cu}/F_2^{D}$ & EMC       &  \cite{EMC}   & 19  & 18.62    \\    
$F_2^{Fe}/F_2^{D}$ & E139      &  \cite{SLAC1} & 23  & 34.95    \\  
$F_2^{Ag}/F_2^{D}$ & E139      &  \cite{SLAC1} & 7   & 9.71     \\ 
$F_2^{Sn}/F_2^{D}$ & EMC       &  \cite{EMC}   & 8   & 16.59    \\ 
$F_2^{Au}/F_2^{D}$ & E139      &  \cite{SLAC1} & 18  & 10.46    \\
$F_2^{C}/F_2^{Li}$ & NMC       &  \cite{NMC1}  & 24  & 33.17    \\  
$F_2^{Ca}/F_2^{Li}$& NMC       &  \cite{NMC1}  & 24  & 25.31    \\  
$F_2^{Be}/F_2^{C}$ & NMC       &  \cite{NMC2}  & 15  & 11.76      \\   
$F_2^{Al}/F_2^{C}$ & NMC       &  \cite{NMC2}  & 15  & 6.93      \\   
$F_2^{Ca}/F_2^{C}$ & NMC       &  \cite{NMC2}  & 15  & 7.71      \\   
$F_2^{Ca}/F_2^{C}$ & NMC       &  \cite{NMC2}  & 24  & 26.09      \\   
$F_2^{Fe}/F_2^{C}$ & NMC       &  \cite{NMC2}  & 15  & 10.38      \\   
$F_2^{Sn}/F_2^{C}$ & NMC       &  \cite{NMC2}  & 15  & 4.69       \\
$F_2^{Sn}/F_2^{C}$ $Q^2$ dep. & NMC   &  \cite{NMC3}  & 145 & 102.31       \\  
$F_2^{Pb}/F_2^{C}$ & NMC       &  \cite{NMC2}  & 15  & 9.57       \\   \hline
$F_2^{\nu Fe}$ & NuTeV       &  \cite{Tzanov:2005kr}  & 78  & 109.65       \\ 
$F_3^{\nu Fe}$ & NuTeV       &  \cite{Tzanov:2005kr}  & 75  & 79.78      \\
$F_2^{\nu Fe}$ & CDHSW       &  \cite{Berge:1989hr} & 120   & 108.20     \\ 
$F_3^{\nu Fe}$ & CDHSW       &  \cite{Berge:1989hr}  & 133  & 90.57      \\ 
$F_2^{\nu Pb}$ & CHORUS      &  \cite{ref:CHORUS}  & 63  &  20.42      \\ 
$F_3^{\nu Pb}$ & CHORUS      &  \cite{ref:CHORUS}  & 63  & 79.58      \\  \hline
$d\sigma^{C}_{DY} /d\sigma^{D}_{DY}$  & E772 & \cite{e772} & 9  & 9.87 \\ 
$d\sigma^{Ca}_{DY}/d\sigma^{D}_{DY}$ & E772 & \cite{e772} & 9  & 5.38 \\
$d\sigma^{Fe}_{DY}/d\sigma^{D}_{DY}$ & E772 & \cite{e772} & 9  & 9.77 \\
$d\sigma^{W}_{DY} /d\sigma^{D}_{DY}$  & E772 & \cite{e772} & 9  & 19.29 \\
$d\sigma^{Fe}_{DY}/d\sigma^{Be}_{DY}$  & E866 & \cite{e866} & 28  & 20.34 \\ 
$d\sigma^{W}_{DY} /d\sigma^{Be}_{DY}$  & E866 & \cite{e866} & 28  & 26.07 \\ \hline
$d\sigma^{dAu}_{\pi^0}/d\sigma^{pp}_{\pi^0}$  & PHENIX & \cite{ref:phenixpi0} & 20 & 27.71 \\  
$d\sigma^{dAu}_{\pi^0}/d\sigma^{pp}_{\pi^0}$  & STAR & \cite{ref:starpi0} & 11 & 3.92 \\
$d\sigma^{dAu}_{\pi^{\pm}}/d\sigma^{pp}_{\pi^{\pm}}$  & STAR & \cite{ref:starpipm} & 30 & 36.63\\ \hline
Total              &           &               & 1579 & 1544.70\\
\end{tabular}
\end{ruledtabular}
\end{table}
In order to obtain DIS structure functions or parton densities for deuterium,
needed, for instance, to compute pion yields in $dAu$ collisions,
we neglect any nuclear effects, assume isospin symmetry ($u^p=d^n$ and $d^p=u^n$),  
and use the free proton PDFs of MSTW \cite{Martin:2009iq}.
Nuclear effects in deuterium were studied in \cite{ref:hirai} by analyzing data
on $F_2^d/F_2^p$ \cite{ref:f2d} and found be small, ${\cal{O}}(1-2\,\%)$, 
in particular, compared to typical uncertainties of nuclear DIS or DY data, 
which justifies our approach of ignoring them.
Parton densities in nuclei with $A>2$ are constructed from the proton densities
bound in a nucleus $A$ as defined in Eq.~(\ref{eq:npdfdef}), assuming that isospin symmetry
also holds for bound systems. 
For instance, the $u$ quark density in a nucleus $A$ with $Z$ protons and
$A-Z$ neutrons at scale $\mu$ is given by
\begin{equation}
\label{eq:updf}
u^A(x_N,\mu) = \frac{Z}{A} f_u^A(x_N,\mu) + \frac{A-Z}{A} f_d^A(x_N,\mu)\;,
\end{equation}
and similarly for $d^A, \bar{u}^A$, and $\bar{d}^A$.
Since most data are given in terms of ratios of structure functions or cross sections,
uncertainties in the free proton PDFs in (\ref{eq:npdfdef}), 
which can be still substantial for less well determined quark flavors or the gluon at 
small $x$ and low scales $Q$ \cite{Martin:2009iq}, are expected to cancel to a large extent.
Neutrino induced DIS off nuclei \cite{Tzanov:2005kr,Berge:1989hr,ref:CHORUS}
is a notable exception since the results are presented as absolute 
structure functions $F_{2,3}^{\nu A}$ instead of ratios.
In order to account for uncertainties in the free proton PDFs, we utilize the
Hessian eigenvector PDF sets of Ref.~\cite{Martin:2009iq} to estimate 
the expected impact of these PDF variations on the results of our fit
to $F_{2,3}^{\nu A}$. We add these additional theoretical uncertainties
in quadrature to the statistical and systematic errors for $F_{2,3}^{\nu A}$.

The total $\chi ^2$ for the optimum fit was found to be 1544.7 for
1579 data points and 25 free fit parameters describing our nPDFs for quarks
and gluons, i.e., a $\chi^2$ per degree of freedom very close to unity
($\chi^2/d.o.f.=0.994$).
In general, all data sets corresponding to different types of observables 
are adequately reproduced, well within the nominal statistical range 
$\chi^2= n \pm \sqrt{2n}$ with $n$ the number of data, cf.\ Tab.~\ref{tab:expchi2}.
More specifically, the partial contribution to $\chi^2$ of all the charged lepton DIS data
amounts to 897.52 units for 894 data points, for neutrino DIS we find
488.20 units compared to 532 data points, DY observables amount to 90.72 units 
for 92 points, and pion production in $dAu$ collisions adds another 68.26 units to $\chi^2$
for 61 data points.
We wish to stress again that there is no need to increase the weight $\omega_i$  
of any particular data set in (\ref{eq:chi2}) to reproduce them well in our global analysis.

\begin{figure}[h]
\begin{center}
\vspace*{-0.6cm}
\epsfig{figure=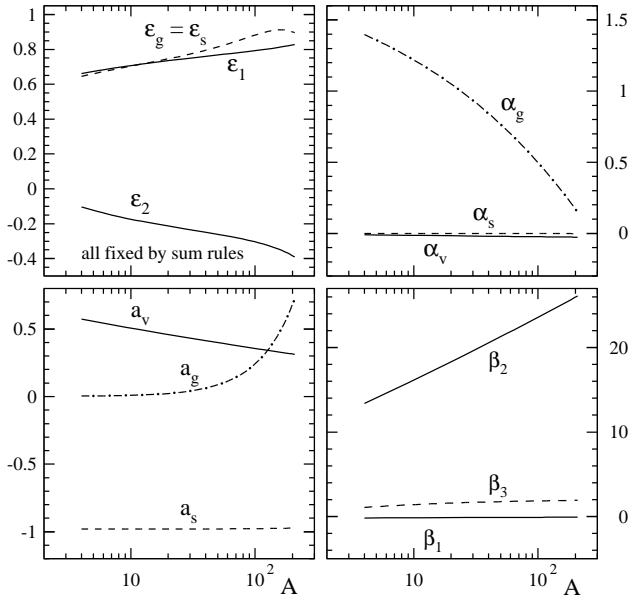,width=0.50\textwidth}
\end{center}
\vspace*{-0.5cm}
\caption{\label{fig:para-adep}
$A$ dependence of the fit parameters according to Tab.~\ref{tab:para}
and Eq.~(\ref{eq:adep}).
Note that $\epsilon_{1,2,g,s}$ are fixed by the sum rules (\ref{eq:charge}),
(\ref{eq:mom}) and assuming $\epsilon_g=\epsilon_s$.}
\end{figure}

The result of the fit suggests that the different data sets are complementary in 
determining the nuclear modifications of PDFs for bound protons and that the 
chosen parametrization in Eqs.~(\ref{eq:rval})-(\ref{eq:rglue})
and (\ref{eq:adep}) is flexible enough to accommodate all the features of the data.
We do not observe any noticeable tension among the different sets of data in the fit.
The parameters describing our optimum set of nPDFs are listed in Tab.~\ref{tab:para}
and their $A$ dependence is illustrated in Fig.~\ref{fig:para-adep}.
We note that the small values for $\lambda_{\xi}$ for some of the parameters
in Tab.~\ref{tab:para} become relevant for large $A$ 
and, hence, are not set to zero.
In Tab.~\ref{tab:eps} we present for completeness the values for $\epsilon_{1,2,s,g}$ 
fixed by charge and momentum conservation, Eqs.~(\ref{eq:charge}) and (\ref{eq:mom}), 
and assuming $\epsilon_s=\epsilon_g$.
%
\begin{table}[th!]
\caption{\label{tab:para} Parameters describing our optimum
NLO ${\overline{\mathrm{MS}}}$ nPDFs in Eqs.~(\ref{eq:rval})-(\ref{eq:rglue})
and (\ref{eq:adep}) at the input scale $Q_0=1\,\mathrm{GeV}$.}
\begin{ruledtabular}
\begin{tabular}{lccc} 
parameter      & $\gamma$ & $\lambda$ & $\delta$  \\ \hline
$\alpha_v$  & -0.256 &0.252 & -0.017 \\         
$\alpha_s$  &0.001 & $-6.89\times10^{-4}$  &  0.286 \\     
$\alpha_g$  &1.994 & -0.401 &  0.286 \\         
$\beta_1$ & -5.564 & 5.36  & 0.0042 \\          
$\beta_2$ & -59.62 & 69.01 & 0.0407 \\          
$\beta_3$ & 2.099 & -1.878 & -0.436 \\          
$a_v$ & -0.622 &1.302 & -0.062 \\               
$a_s$ & -0.980 & $2.33\times 10^{-6}$ & 1.505 \\           
$a_g$ & 0.0018 & $2.35 \times 10^{-4}$ & 1.505 \\          
\end{tabular}
\end{ruledtabular}
\end{table}
%
\begin{table}[ht!]
\caption{\label{tab:eps} Values for $\epsilon_{1,2,s,g}$ for our optimum fit in Tab.~\ref{tab:para} 
for selected nuclei A 
as obtained from charge and momentum conservation, Eqs.~(\ref{eq:charge}) and (\ref{eq:mom}),
and assuming $\epsilon_s=\epsilon_g$.}
\begin{ruledtabular}
\begin{tabular}{cccc} 
   A  &  $\epsilon_1$  &  $\epsilon_2$  &  $\epsilon_s=\epsilon_g$     \\ \hline  
   4  &  0.6612      &   -0.1033    &    0.6448     \\
  12  &  0.7149      &   -0.1851    &    0.7147     \\
  27  &  0.7458      &   -0.2287    &    0.7655     \\
  40  &  0.7596      &   -0.2487    &    0.7947     \\
  56  &  0.7714      &   -0.2668    &    0.8239     \\
 197  &  0.8245      &   -0.3811    &    0.9020     \\
 208  &  0.8280      &   -0.3912    &    0.8952     \\
\end{tabular}
\end{ruledtabular}
\end{table}
%
%

\subsection{Charge lepton DIS and DY data \label{sec:disdy}}
%
We continue the discussion of the results of our fit with
a detailed comparison with the available charged lepton DIS and
DY data, which are the core part of all extractions of nPDFs 
\cite{ref:nds,ref:hirai,ref:eps09,ref:schienbein,Kovarik:2010uv}.

\begin{figure}[th!]
\begin{center}
\vspace*{-0.6cm}
\epsfig{figure=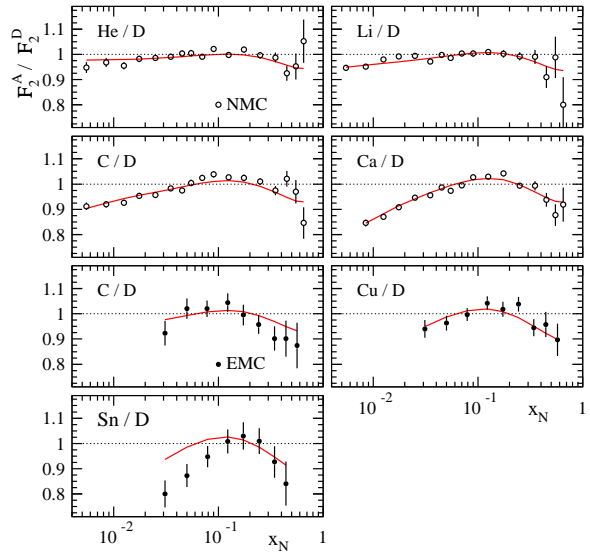,width=0.48\textwidth}
\end{center}
\vspace*{-0.5cm}
\caption{\label{fig:disdata1}
Data for the DIS structure function ratio $F_2^A/F_2^D$ from EMC \cite{EMC} and NMC
\cite{NMC1} as a function of momentum fraction $x_N$ compared with the result of
our global fit.}
\end{figure}

\begin{figure}[h!]
\begin{center}
\vspace*{-0.6cm}
\epsfig{figure=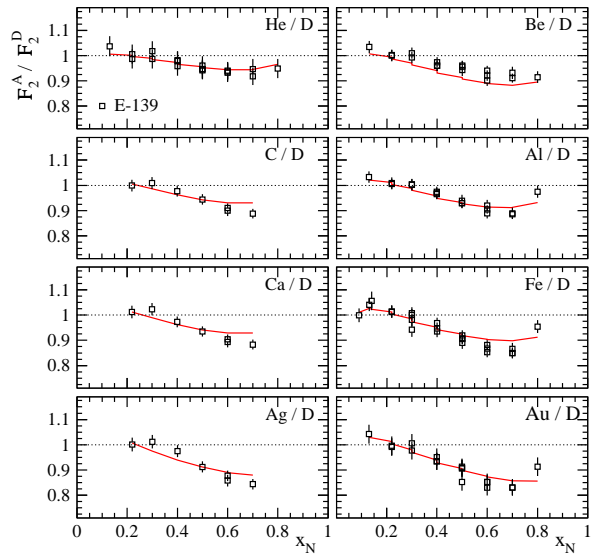,width=0.48\textwidth}
\end{center}
\vspace*{-0.5cm}
\caption{\label{fig:disdata2}
The same as in Fig.~\ref{fig:disdata1} but now for the SLAC E-139 data \cite{SLAC1}.
Note that the multiple points for a given $x_N$ have different
$Q^2$ values in the range $2-10\,\mathrm{GeV}^2$.}
\end{figure}

Figures~\ref{fig:disdata1} and \ref{fig:disdata2} show the ratios $F_2^A/F_2^D$
of the DIS structure functions for various nuclei $A$
with respect to deuterium from EMC and NMC \cite{EMC,NMC1} 
and the E-139 collaboration \cite{SLAC1}, respectively.
The solid lines correspond to the result of the fit 
at the scale $Q^2$ of each data point, where
$F^A_2$ is computed with our NLO set of nPDFs and
$F_2^D$ is obtained with the free proton PDFs of \cite{Martin:2009iq}.
%
%
%
%
%
%
\begin{figure}[h!]
\begin{center}
\vspace*{-0.6cm}
\epsfig{figure=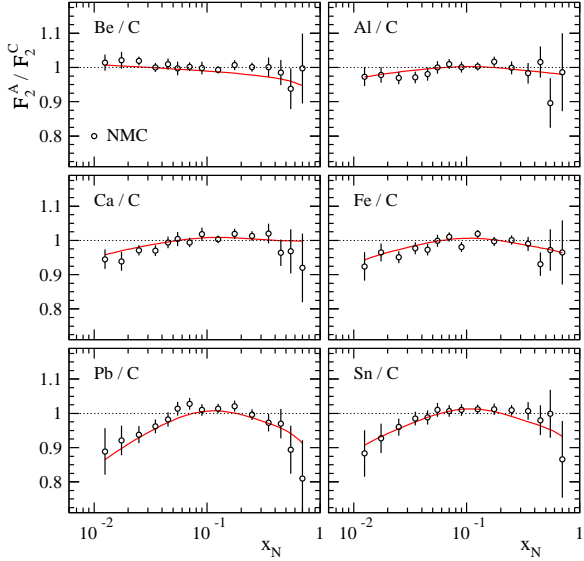,width=0.48\textwidth}
\end{center}
\vspace*{-0.5cm}
\caption{\label{fig:disdata3}
The same as in Fig.~\ref{fig:disdata1} but now
for $F_2^A/F_2^C$ from NMC \cite{NMC2}.}
\end{figure}
Similarly, Figs.~\ref{fig:disdata3} and \ref{fig:disdata4} show comparisons with 
structure function ratios using carbon and lithium as reference \cite{NMC1,NMC2}.
The data clearly show the well known regions of shadowing, anti-shadowing, and large $x_N$ EMC effect for
$x_N\lesssim 0.05$, $x_N\approx 0.1$, and $x_N\gtrsim 0.3$, respectively, and are in general
well reproduced by the fit at NLO accuracy, cf.~Tab.~\ref{tab:expchi2} for individual $\chi^2$ values.
The only exception is the low $x_N$ behavior of $F_2^{Sn}/F_2^D$ 
in Fig.~\ref{fig:disdata1}, however, the fit reproduces very well both the ratio $F_2^{Sn}/F_2^C$ in Fig.~\ref{fig:disdata2} and $F_2^{C}/F_2^D$ in Fig.~\ref{fig:disdata1}.
We notice, that the strong rise of the ratios to values larger than unity as $x_N\to 1$ due to Fermi motion 
is not seen in the analyzed DIS data but shows 
up prominently in low $Q^2$ experiments, see, e.g.~\cite{Seely:2009gt}, where target mass corrections
are relevant though.

%
\begin{figure}[th!]
\begin{center}
\vspace*{-0.6cm}
\epsfig{figure=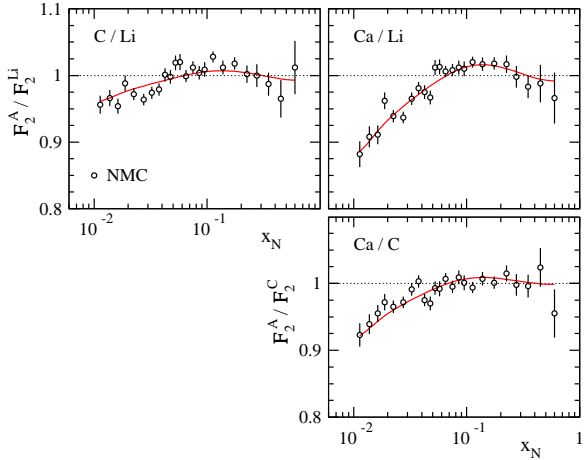,width=0.48\textwidth}
\end{center}
\vspace*{-0.5cm}
\caption{\label{fig:disdata4}
The same as in Fig.~\ref{fig:disdata1} but now
for $F_2^A/F_2^{Li}$ \cite{NMC1} and $F_2^A/F_2^C$ \cite{NMC2} from NMC.}
\end{figure}
%
%
\begin{figure}[h!]
\begin{center}
\vspace*{-0.6cm}
\epsfig{figure=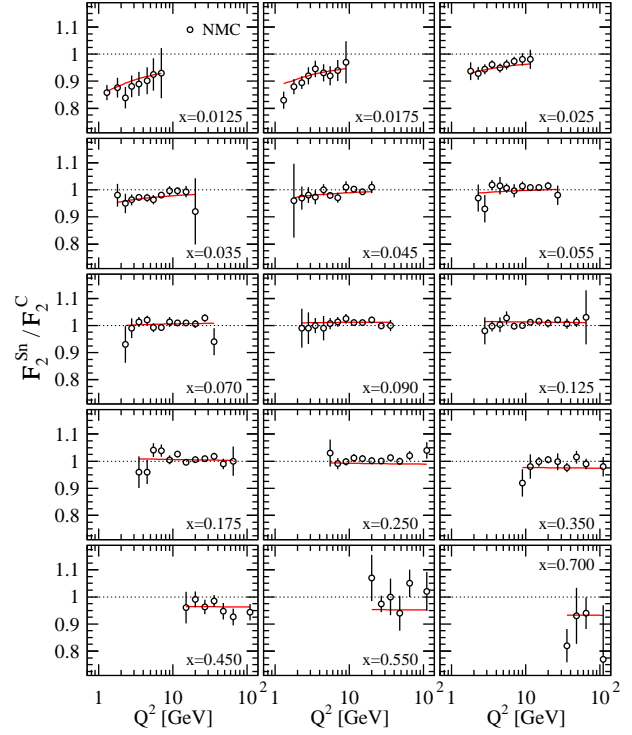,width=0.499\textwidth}
\end{center}
\vspace*{-0.5cm}
\caption{\label{fig:disdata5}
Data on the $Q^2$ dependence of ratio $F_2^{Sn}/F_2^C$ for fixed bins in $x_N$
from NMC \cite{NMC3} compared to the result of our global fit.}
\end{figure}
%
%
%
\begin{figure}[th!]
\begin{center}
\vspace*{-0.6cm}
\epsfig{figure=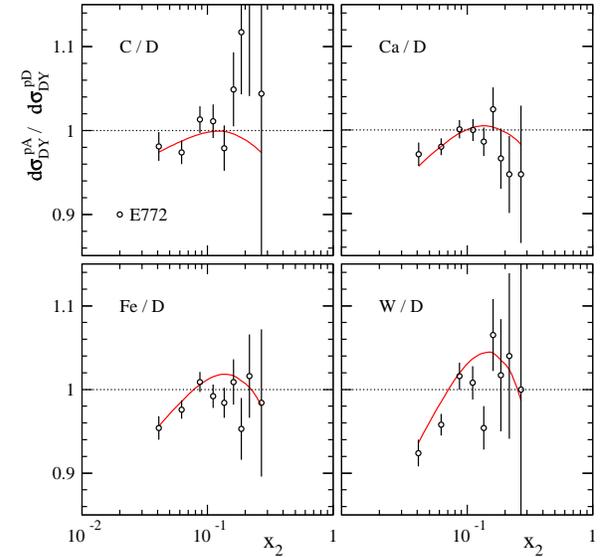,width=0.48\textwidth}
\end{center}
\vspace*{-0.5cm}
\caption{\label{fig:dydata1}
Ratios of nuclear DY di-muon yields with invariant mass $M\ge 4\,\mathrm{GeV}$ 
from E772 \cite{e772} as a function of the parton momentum fraction $x_2$ of the nucleus. 
The solid lines are the result of our fit.}
\end{figure}
%
%
\begin{figure}[h!]
\begin{center}
\vspace*{-0.6cm}
\epsfig{figure=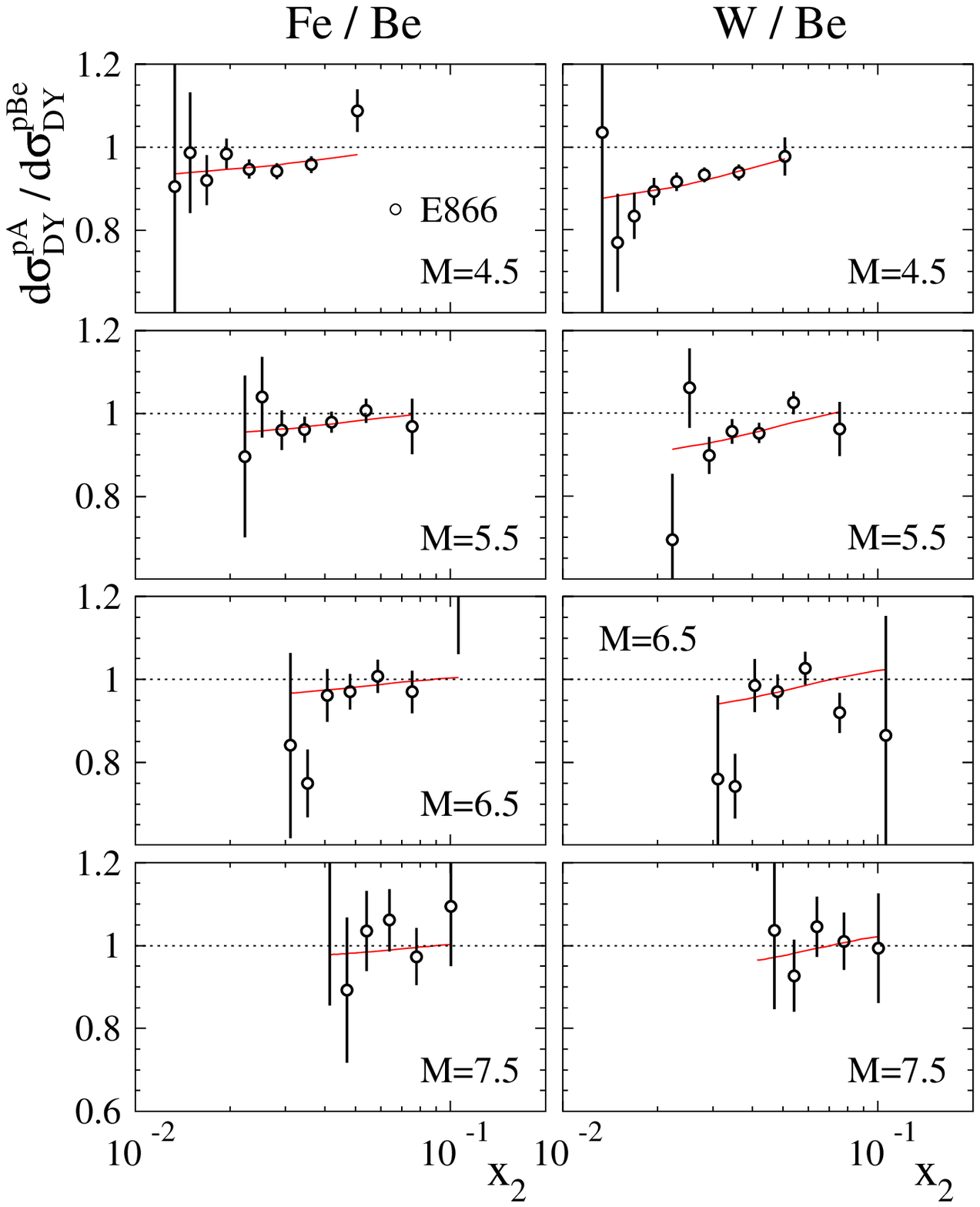,width=0.48\textwidth}
\end{center}
\vspace*{-0.5cm}
\caption{\label{fig:dydata2}
Similar as in Fig.~\ref{fig:dydata1} but now for the E866 data \cite{e866} 
in various bins of the invariant mass $M$ of the di-muon pair.}
\end{figure}
For the imposed cut $Q^2>1\,\mathrm{GeV}^2$, the analyzed charged lepton DIS data cover the range $x_N\gtrsim 0.01$ 
and about a decade in $Q^2$ for any given bin in $x_N$ as is illustrated in Fig.~\ref{fig:disdata5}
for the ratio $F_2^{Sn}/F_2^C$ \cite{NMC3}. 
The observed, rather moderate $Q^2$ dependence, best visible for the smallest $x_N$, 
provides some constraint on the nuclear modifications of the gluon density through DIS scaling violations.
Our fit compares well with similar data on the $Q^2$ dependence of the ratios for $F_2^{Li}/F_2^{D}$ and
$F_2^{C}/F_2^{D}$ \cite{NMC1}, see Tab.~\ref{tab:expchi2}.
Not surprisingly, no substantial differences to previous NLO analyses of nPDFs, such as \cite{ref:nds}
or \cite{ref:eps09}, are found in the quality of the description of charged lepton nuclear DIS data.

%
%
\begin{figure*}[tbh!]
\begin{center}
\vspace*{-0.6cm}
\epsfig{figure=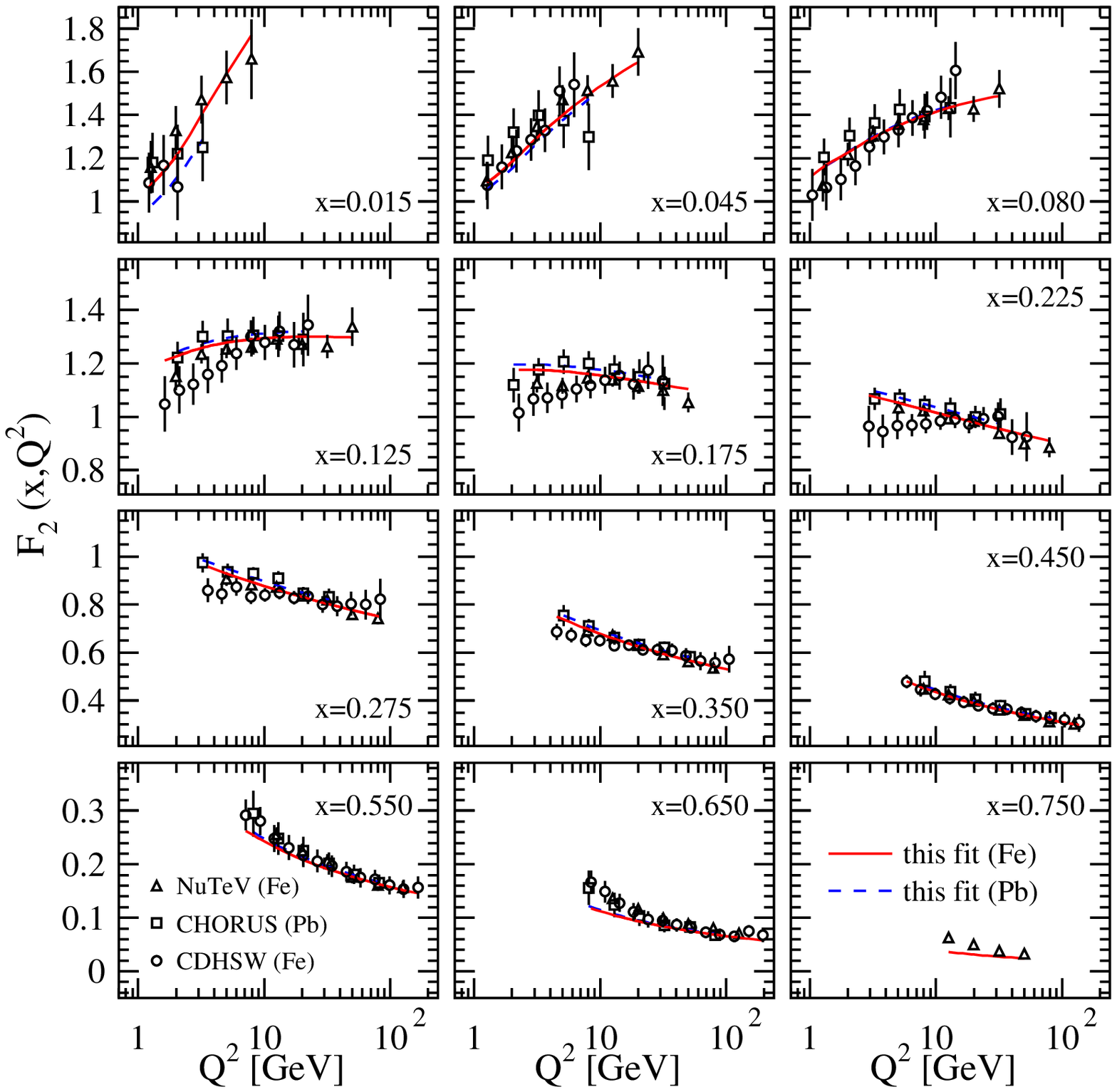,width=0.7\textwidth}
\end{center}
\vspace*{-0.5cm}
\caption{\label{fig:f2neutrino}
The averaged CC DIS structure function $F_2$ as a function of $Q^2$
in various bins of $x_N$ for iron \cite{Tzanov:2005kr,Berge:1989hr}
and lead \cite{ref:CHORUS} targets compared with the result of our NLO fit
shown as solid and dashed lines, respectively. }
\end{figure*}
Although a distinction of nuclear modifications for quarks and antiquarks is not possible 
based on the charged lepton DIS data alone, they provide a valuable constraint on nPDFs by
mainly probing $R_v$ at medium-to-large $x_N$ and $R_s$ at the lowest available momentum 
fractions $x_N\simeq 0.01$. 
Despite being not too accurate, DY di-muon production data \cite{e772,e866} obtained in $pA$ collisions 
help to disentangle valence and sea quarks further.
Again, experimental results are presented as ratios of 
$pA$ cross sections, see Eq.~(\ref{eq:dyxsec}),  
for heavier nuclei and either deuterium \cite{e772} or 
beryllium \cite{e866} targets and are shown in Figs.~\ref{fig:dydata1} and \ref{fig:dydata2},
respectively.  

The E772 and E866 DY data, obtained with an $800\,\mathrm{GeV}$ proton beam
incident on a fixed target and for invariant masses $M>4\,\mathrm{GeV}$
of the di-muon pair, probe the range $0.01\lesssim x_2 \lesssim 0.2$
of momentum fractions $x_2$ in the heavy nuclei. 
Ratios smaller than unity can be taken as an indication of shadowing for sea quark densities
in nuclei, i.e., $R_s<1$ at small $x_N$.
The relatively large scale of the data, set by the invariant mass $M$ of the di-muon pair,
provides some handle on evolution effects in the global fit.

\subsection{Neutrino induced DIS off nuclear targets}
%
Charged current (CC) neutrino DIS data \cite{Tzanov:2005kr,Berge:1989hr,ref:CHORUS} are one of the major additions 
to our previous analysis in \cite{ref:nds} and are subject to an ongoing 
discussion \cite{ref:schienbein,Kovarik:2010uv,ref:paukkunen} 
about their compatibility, or lack thereof, with the neutral current (NC) DIS data discussed
in Subsec.~\ref{sec:disdy}.
A good understanding of potential issues with neutrino DIS data is also relevant
for conventional PDF analyses of free protons where they provide a vital constraint
on the strangeness distribution.

%
\begin{figure*}[tbh!]
\begin{center}
\vspace*{-0.6cm}
\epsfig{figure=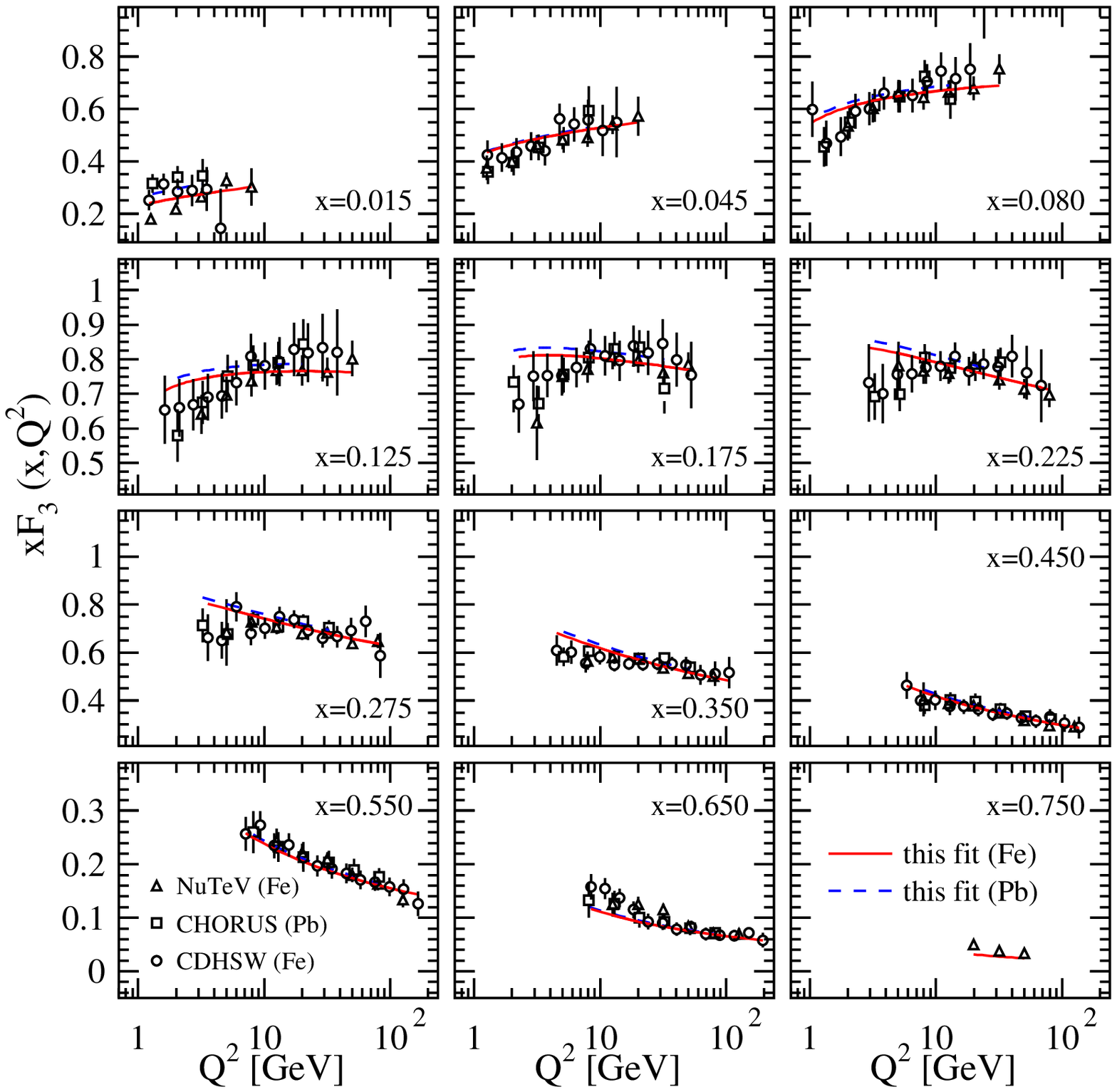,width=0.7\textwidth}
\end{center}
\vspace*{-0.5cm}
\caption{\label{fig:xf3neutrino}
Same as in Fig.~\ref{fig:f2neutrino} but now for the averaged CC structure function $x_N F_3$. }
\end{figure*}
Neglecting complications due to Cabibbo-Kobayashi-Maskawa mixing for the sake of argument,
CC DIS data draw their relevance for global PDF fits from the different combinations of
up-type and down-type quark flavors they are sensitive to. 
With neutrino and antineutrino beams one can probe
four different structure functions in CC DIS off a nucleon $A$ given, to LO accuracy, by
\begin{eqnarray}
\nonumber
F_2^{\nu A}(x_N) &\simeq& x_N [\bar{u}^A + \bar{c}^A + d^A + s^A]\,(x_N)\;, \\[2mm]
\nonumber
F_2^{\bar{\nu} A} (x_N) &\simeq& x_N [u^A + c^A + \bar{d}^A + \bar{s}^A]\,(x_N)\;, \\[2mm]
\nonumber
F_3^{\nu A} (x_N) &\simeq& [ -(\bar{u}^A + \bar{c}^A) + d^A + s^A]\,(x_N)\;,  \\[2mm]
F_3^{\bar{\nu} A} (x_N) &\simeq& [ u^A + c^A - (\bar{d}^A + \bar{s}^A)]\,(x_N)\;,
\label{eq:nudis}
\end{eqnarray}
where we have suppressed the scale dependence. 
Assuming, as usual, that isospin symmetry holds to a good approximation
for bound protons and neutrons, the $u^A$ density in a nucleus $A$ in (\ref{eq:nudis})
is given by Eq.~(\ref{eq:updf}) and similarly for $d^A, \bar{u}^A$, and $\bar{d}^A$.
Experiments extract, under certain assumptions, 
averaged structure functions $F_{2,3}\equiv (F_{2,3}^{\nu A} + F_{2,3}^{\bar{\nu}A})/2$
from appropriate linear combinations of neutrino and antineutrino CC DIS differential
cross sections \cite{Tzanov:2005kr,Berge:1989hr,ref:CHORUS}, corrected for QED radiative corrections.
As can be easily inferred from (\ref{eq:nudis}), $F_2$ is proportional to the total singlet
combination of quarks and antiquarks and hence sensitive to both valence and sea quarks depending
on the value of $x_N$. Since the kinematic coverage of CC and NC $F_2$ data overlaps to some extent,
any significant tension between these two measurements should show up prominently in a
global QCD analysis.
Any different nature of interactions of photons and charged weak bosons with nuclear matter
shall result in different, non-universal sets of nPDFs.
Likewise, the averaged CC structure function $F_3$ mainly probes the valence combination 
$u_v^A+d_v^A$, which is already well constrained at sufficiently large $x_N$
by the NC data discussed in Subsec.~\ref{sec:disdy}.
By combining CC $F_3$ and NC $F_2$ data one can arrive at a much improved valence and sea
quark separation in the entire $x_N$ region where data overlap.

As it turns out, the CC data for the averaged structure function $F_2$ are
remarkably well reproduced within the experimental uncertainties 
by our fit, both in shape and in magnitude as is illustrated by Fig.~\ref{fig:f2neutrino}.
The only noticeable exception are the CDHSW data \cite{Berge:1989hr}
at $Q^2$ values below $10\,\mathrm{GeV}^2$ where they exhibit a rather different slope 
than the other data. In fact, in this $Q^2$ region it appears
to be impossible to simultaneously fit all data sets equally well, suggesting some
systematic discrepancy among the different neutrino data which needs to be further investigated.
Data for the averaged structure function $x_N F_3$ are also well described by our fit as
can be seen in Fig.~\ref{fig:xf3neutrino}, except perhaps for the lowest $Q^2$ values 
in some bins at intermediate $x_N$ where the slopes do not match.
Both sets of data, $F_2$ and $x_N F_3$, feature the typical pattern of scaling violations,
i.e., they increase and decrease with $Q^2$ for $x_N\lesssim 0.2$ and $x_N\gtrsim 0.2$,
respectively.
As mentioned above, the error bars in Figs.~\ref{fig:f2neutrino} and \ref{fig:xf3neutrino}
comprise the statistical and systematic uncertainties of the data added in quadrature and
estimates of theoretical ambiguities from variations of the PDFs of free nucleons which 
are most relevant at small $x_N$ and low $Q^2$ \cite{Martin:2009iq}.
The latter are included because neutrino induced DIS data are presented as absolute
cross section or structure function measurements rather than ratios
in the absence of $\nu p$ or $\nu D$ benchmark results. 
Extractions of the nuclear modification factors $R_i^A$ in (\ref{eq:npdfdef}) from CC data
are hence more sensitive to the assumptions about the PDFs of free nucleons and their
uncertainties.

%
\begin{figure}[th!]
\begin{center}
\vspace*{-0.2cm}
\epsfig{figure=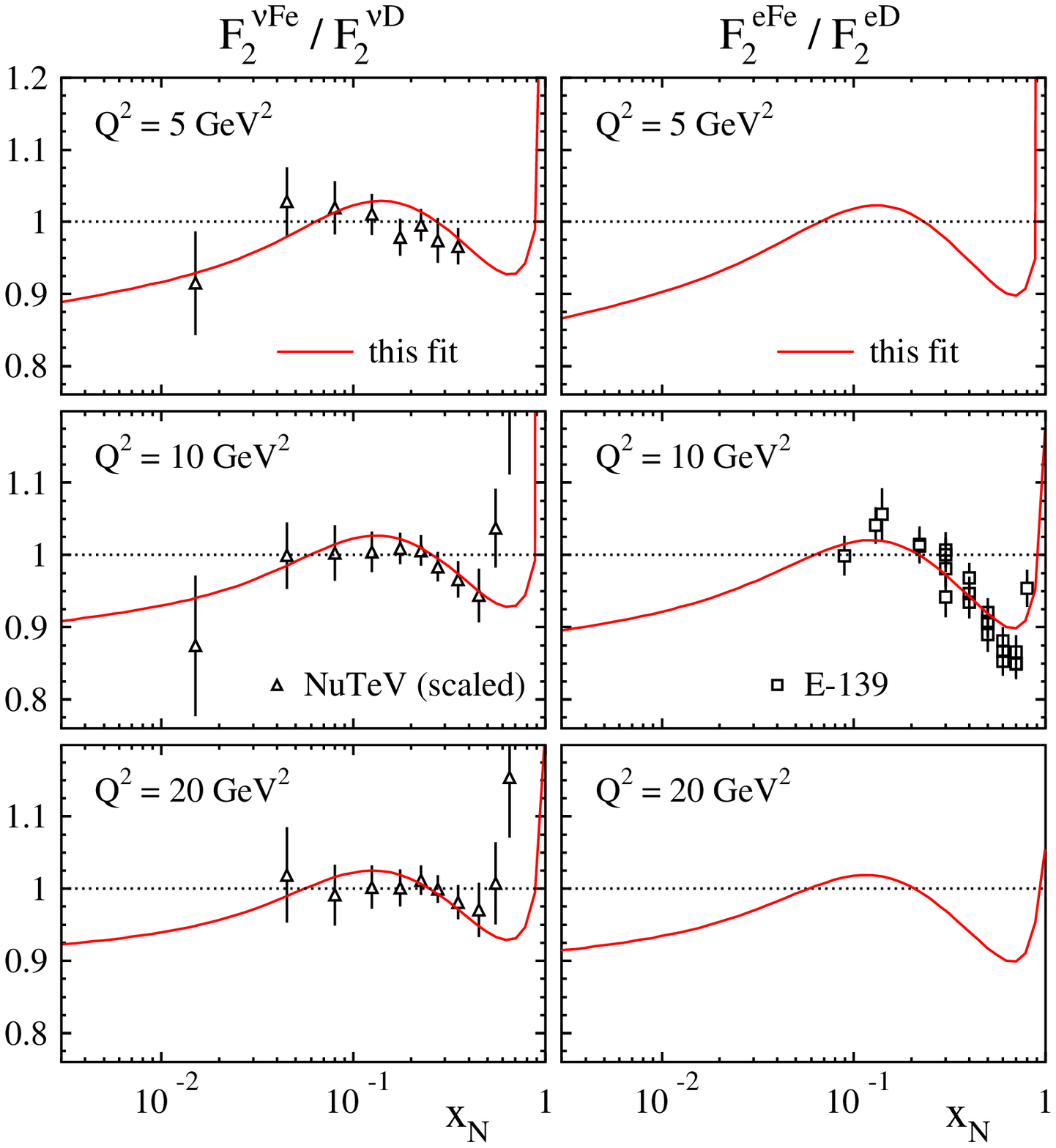,width=0.48\textwidth}
\end{center}
\vspace*{-0.5cm}
\caption{\label{fig:neutrinoratios}
{\bf left:} data and theory driven estimates of the ratio $F_2^{\nu Fe}/F_2^{\nu D}$,
see text for details. {\bf right:} nuclear modifications from charged lepton DIS,
$F_2^{eFe}/F_2^{eD}$, as obtained from the fit to SLAC E-139 data.} 
\end{figure}
At variance with our results, a significant tension between CC and NC current
nuclear DIS data was reported in Refs.~\cite{ref:schienbein,Kovarik:2010uv}
based on a fit of the several thousand data points on differential 
$\nu A$ and $\bar{\nu}A$ cross sections rather than the averaged CC structure
functions  $F_{2,3}$ used in our analysis.
Their result, if true, casts serious doubt on the validity of pQCD factorization
for processes involving bound nucleons as it suggests a different, non universal
behavior of nuclear corrections $R_i^A$ in CC neutrino induced and NC charged lepton DIS.
This is illustrated, e.g., in Fig.~1 of \cite{Kovarik:2010uv} where the authors
compute the ratio $F_2^{\nu Fe}/F_2^{\nu D}$ from the NuTeV data 
with an iron target and an estimate of
the hypothetical structure function $F_2^{\nu D}$ for neutrino DIS off deuterium,
which differs significantly from the measured $F_2^{l Fe}/F_2^{l D}$
obtained from charged lepton DIS data.

However, our global QCD analysis of nPDFs, in particular, the results presented in 
Figs.~\ref{fig:f2neutrino} and \ref{fig:xf3neutrino} and Tab.~\ref{tab:expchi2}, do not support 
such a strong conclusion which would have far reaching implications not only for extractions 
of nPDFs but for free proton PDFs as well as $\nu A$ DIS data for isoscalar targets 
are often used to constrain the strangeness and anti-strangeness densities.
We notice that also in Ref.~\cite{ref:paukkunen} no apparent disagreement between the nuclear modification factors
$R_i^A$ for CC and NC DIS data has been found based on a comparison of an existing fit \cite{ref:eps09},
not including CC data, with the same set of data points for $\nu A$ and $\bar{\nu}A$ cross sections
used in \cite{ref:schienbein,Kovarik:2010uv}.

To further illustrate the consistent picture of nuclear modifications emerging from 
our fit, we also show estimates of the ratio $F_2^{\nu Fe}/F_2^{\nu D}$ 
on the left hand side of Fig.~\ref{fig:neutrinoratios}. 
The experimental results are obtained with the NuTeV data closest to
the $Q^2$ values selected in the plot, rescaled by a NLO 
calculation of $F_2^{\nu D}$ adopting our reference set free proton PDFs 
from MSTW \cite{Martin:2009iq}, including heavy quark mass effects, and
assuming that nuclear effects are negligible for deuterium.
These data-based ratios exhibit a pattern of nuclear modifications which 
is not exactly the typical one but resembles all of the expected features in
that it has shadowing, anti-shadowing, EMC, and Fermi motion effects of similar magnitude,
at low, intermediate, and large values of $x_N$, respectively.
The solid lines are computed in a similar way but now using the result
of our fit to the NuTeV data. These fit-driven ratios 
are clearly consistent with the data-based $F_2^{\nu Fe}/F_2^{\nu D}$ 
within the present, rather large uncertainties, except for the largest values
of $x_N$, $x_N=0.65$ and $0.75$, where the theoretical estimate of
$F_2^{\nu D}$ is most likely not reliable enough and where  
target mass corrections might become relevant \cite{ref:paukkunen}.
We believe that the way in which $F_2^{\nu D}$ is estimated is the main difference with
respect to what is shown in \cite{ref:schienbein,Kovarik:2010uv}.
On the right hand side of Fig.~\ref{fig:neutrinoratios} we present for comparison
the corresponding results for the nuclear modifications $F_2^{e Fe}/F_2^{e D}$ 
obtained from DIS of charged leptons off an iron target. 
With the exception of a slightly more pronounced dip from the EMC effect at
large $x_N$, the ratios for CC and NC DIS are very similar and, unlike 
Fig.~1 in Ref.~\cite{Kovarik:2010uv}, no significant tension is observed.
A moderate difference between the two ratios should be actually expected as
they probe different combinations of quark densities. However, a 
flexible enough parametrization of nuclear effects $R_i^A$ can accommodate all sets 
of data equally well.

We close the discussion on CC DIS by noticing that
the proper treatment of heavy quark mass effects is an important asset
of our global analysis. The mass dependence is fully accounted for by using the 
recently obtained expressions of the NLO coefficients \cite{ref:nlocc} 
in Mellin moment space \cite{Blumlein:2011zu}. These corrections are
known to be of particular relevance for the strangeness contribution to CC DIS 
which produces a massive charm quark in the final state, and they have a positive impact
on the quality of the fit in terms of $\chi^2$.

\subsection{Pion production in $dAu$ collisions \label{sec:dau}}
%
%
\begin{figure}[th!]
\begin{center}
\vspace*{-0.6cm}
\epsfig{figure=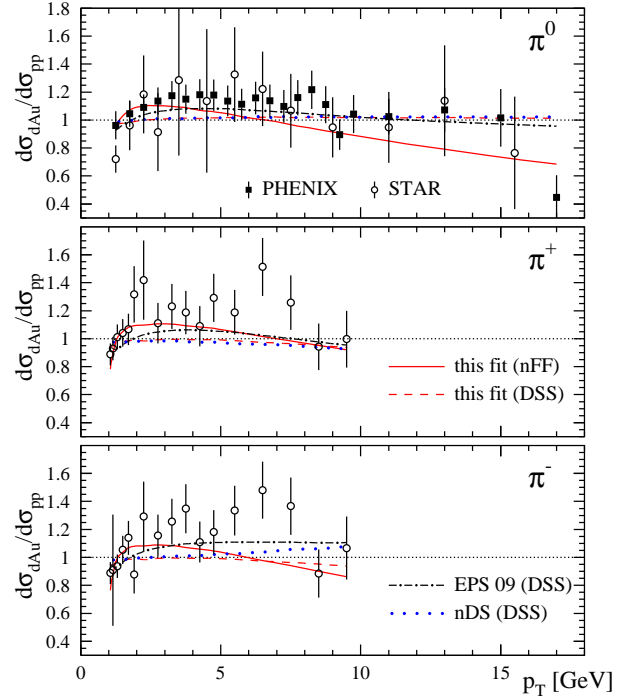,width=0.48\textwidth}
\end{center}
\vspace*{-0.5cm}
\caption{\label{fig:daudata} Ratios of the $dAu$ and $pp$ pion production cross sections per nucleon
at mid rapidity from the PHENIX \cite{ref:phenixpi0} and STAR \cite{ref:starpi0,ref:starpipm} 
collaborations compared to the result of our fit
using modified \cite{ref:nff} (solid lines) or vacuum \cite{ref:dsspion} (dashed lines) FFs. 
Also shown are calculations using the sets of nPDFs from \cite{ref:nds} and \cite{ref:eps09},
dotted and dot-dashed lines, repectively.}
\end{figure}
Data for single inclusive pion production at mid rapidity and high transverse momentum $p_T$ in $dAu$
collisions at RHIC are the other major addition to our previous analysis \cite{ref:nds}.
Figure~\ref{fig:daudata} shows the neutral and charged pion minimum bias production cross sections per nucleon 
for $dAu$ collisions measured by PHENIX \cite{ref:phenixpi0} and STAR \cite{ref:starpi0,ref:starpipm},
normalized to the corresponding yields in $pp$. The ratios are obtained for pions at mid rapidity 
and presented as a function of their $p_T$, which also sets the hard scale for perturbative calculations
using Eq.~(\ref{eq:daxsec}). The various theoretical curves shown in Fig.~\ref{fig:daudata}
are explained and discussed below.

We are limited to using minimum bias data as collinear nPDFs do not exhibit any information on 
the distribution of partons in the transverse plane needed for computations of the impact parameter
or centrality dependence of heavy-ion cross sections. 
Comparing ratios of measured minimum bias $dAu$ and $pp$ cross sections avoids model dependent 
estimates of the average number of binary nucleon-nucleon collisions $\langle N_{coll} \rangle$ in a given 
centrality class, see, e.g., Ref.~\cite{ref:starpi0} for experimental details.
Notice that even in the absence of nuclear effects, these ratios are not necessarily expected
to be unity as they can be affected by isospin effects  such as the smaller density of
$u$ quarks in a nucleus than in a free proton due to the dilution from neutrons. 
However, noticeable numerical effects are only expected for electromagnetic probes like
prompt photons \cite{Arleo:2011gc} which couple directly to the electric charge of the quarks, 
see Sec.~\ref{sec:future}.

In general, results from $dAu$ collisions are significantly less straightforward to 
interpret in terms of nuclear modification factors $R_i^A$ than DIS data. 
Each value of $p_T$ samples different fractions of the contributing
partonic hard scattering processes, integrated over a large range of momentum
fractions $x_N$, and convoluted with information on the PDFs of the deuterium. 
Furthermore, since $p_T$ sets the magnitude for the factorization scale in (\ref{eq:daxsec}), 
the ratios in Figure~\ref{fig:daudata} not only reflect
the amount of nuclear modifications but also their energy scale dependence. 
The presence of hadronic probes such as pions, charmed mesons, or jets
in the final-state, all originating from strongly interacting quarks or gluons, 
leads to additional complications.
Apart from the nuclear effects on parton densities, accounted for by the nPDFs,
the cross sections are in principle also sensitive to medium induced modifications
in the hadronization process.
In case of inclusive hadron production and assuming factorizability of such medium effects,
they should be absorbed into effective nuclear parton-to-hadron 
fragmentation functions (nFFs), denoted as $D_k^{A,\pi}$ in Eq.~(\ref{eq:daxsec}).
Modifications of hadron yields have been found to be quite significant,
with attenuation effects of up to $50\%$ for heavy nuclei,   
for hadron multiplicities in nuclear DIS by the HERMES collaboration \cite{Airapetian:2009jy}.
They have been parametrized in terms of nFFs in Ref.~\cite{ref:nff} in a 
NLO analysis assuming factorization and based on
the HERMES \cite{Airapetian:2009jy} DIS and the RHIC $dAu$ pion production 
\cite{ref:phenixpi0,ref:starpi0,ref:starpipm} data. 

The solid lines in Fig.~\ref{fig:daudata} represent the result of 
our best fit of nPDFs using the nFFs of Ref.~\cite{ref:nff} in the calculation
of the $dAu$ cross section in (\ref{eq:daxsec}) and the standard vacuum FFs 
from the global analysis of DSS \cite{ref:dsspion} to estimate the $pp$ yields.
The fit follows well the rise and fall of the ratio at small and high $p_T$,
respectively, but falls somewhat short in reproducing the enhancement found at medium $p_T$.
Owing to the large experimental uncertainties, the $\chi^2$ for this subset of data is
nevertheless very good, $\chi^2_{dAu}/\#\mathrm{data}=1.12$ and slightly better than
the outcome of an otherwise similar fit using vacuum FFs (dashed lines) 
where $\chi^2_{dAu}/\#\mathrm{data}=1.37$ and $\chi^2_{\mathrm{tot}}=1560.5$.
As it turns out, the use of either nFFs or vacuum FFs in the fit has some
impact on the shape of the different estimates for the ratios in Fig.~\ref{fig:daudata}.
In any case, the obtained nPDFs are contingent upon the accuracy of the set
of FFs used in the analysis and the validity of factorization for pion production in $dAu$ collisions,
making DY di-leptons or prompt photons the much cleaner but experimentally more demanding probe for nPDFs.

Data for neutral pion yields in $dAu$ collisions were first incorporated in the global
analysis by EPS \cite{ref:eps09} and found to provide a vital new constraint on $R_g^{Au}$.
At variance with our approach, the authors in \cite{ref:eps09} disregard any 
medium modifications in the hadronization and, most importantly, put a large
weight $\omega_{dAu}=20$ on this subset of data in the minimization of the
$\chi^2$ function (\ref{eq:chi2}) to maximize its impact.
They achieve a good description of the data as can be seen from the dot-dashed lines
in Fig.~\ref{fig:daudata}.
It is not too surprising that the $dAu$ data 
drive rather pronounced modifications of the gluon nPDF in their fit,
see Fig.~3 in \cite{ref:eps09}, not found in our analysis which uses $\omega_{dAu}=1$. 

In order to better understand the correlation between the $dAu$ data and a potentially
sizable modification of the gluon density, it is instructive to estimate first the mean
momentum fraction $\langle x_N\rangle$ probed at a given $p_T$.
The standard way of obtaining $\langle x_N\rangle$ is to evaluate the convolutions
in (\ref{eq:daxsec}) with an additional factor on $x_N$ in the integrand and then divide
by the cross section itself, see, e.g. Eq.~(4) in \cite{Sassot:2010bh}.
Typically, the $\langle x_N \rangle$ for pion production at mid rapidity at RHIC rises from
about  $0.05$ at $p_T\simeq 1\,\mathrm{GeV}$ 
to around $0.3$ at $p_T\simeq 15\,\mathrm{GeV}$. 
Therefore, for $p_T \gtrsim 1\,\mathrm{GeV}$ the cross section 
is mainly sensitive to nPDFs in the anti-shadowing region and at larger $p_T$ to 
the suppression due to the EMC effect.

A better description of the data at intermediate $p_T$ in \cite{ref:eps09},
where we fall somewhat short, is achieved by an extraordinarily large enhancement (anti-shadowing) of
gluons at $x_N\simeq 0.1$, where quarks are well constrained by DIS and DY data. This is induced by
the large weight $\omega_{dAu}=20$. In our fit, some part of the enhancement in this $p_T$ region is provided
by a slightly larger gluon-to-pion FFs in a nuclear medium, i.e.,  $D_g^{A,\pi}/D_g^{\pi} >1$ \cite{ref:nff}.
At larger values of $p_T$, the region of the EMC effect comes into play,
where again the quarks are already well constrained, and the drop of the data is
reproduced by an $R_g^{Au}$ much less the unity at $x_N\simeq 0.6$.
The smaller quark-to-pion FFs in a medium, $D_q^{A,\pi}/D_q^{\pi} <1$ \cite{ref:nff}, in accordance
with the large hadron attenuation found by HERMES \cite{Airapetian:2009jy},
contributes to the behavior of the ratio found in our fit.
In both regions of $x_N$, the obtained modifications of the gluon nPDF in the EPS
fit are much more pronounced than the corresponding ones for valence or sea quarks
and not supported by our analysis, see Fig.~\ref{fig:rpdf-at10} below, based on $\omega_{dAu}=1$.
 
Finally, we refrain from using $dAu$ data obtained at forward rapidity
by BRAHMS \cite{Arsene:2004ux} and STAR \cite{Adams:2006uz} 
which show a rapidly increasing suppression of the cross section
ratios for $p_T < 2\,\mathrm{GeV}$. Although these data might help
to further constrain the gluon nPDF down to smaller values of $x_N$, the
theoretical scale uncertainties are very large, and the application of pQCD
is questionable. Also, little is known about FFs and a possible medium modification
in this kinematic region.

\subsection{nPDFs and their uncertainties \label{sec:npdf}}
%
Estimating the uncertainties of PDFs and FFs obtained from global $\chi^2$
optimizations has become an important objective.
The most common and practicable approach, the ``Hessian method'', 
explores the uncertainties associated with
the fit through a Taylor expansion of $\chi^2(\{\xi\})$ around 
the global minimum $\chi^2_0(\{\xi_0\})$, where $\{\xi\}$ denotes the set
of free parameters of the chosen functional form at the initial scale $Q_0$
and $\{\xi_0\}$ their values for the optimum fit.
Keeping only the leading quadratic terms, the increase $\Delta \chi^2$
can be written in terms of the Hessian matrix
\begin{equation}
\label{eq:hij}
H_{ij} \equiv \frac{1}{2} \frac{\partial^2 \chi^2}{\partial y_i \partial y_j}
\Bigg|_{0} 
\end{equation}
as
\begin{equation}
\label{eq:chi2hessian}
\Delta \chi^2 =  \chi^2(\{\xi\}) - \chi^2_0(\{\xi_0\}) = 
\sum_{ij}^{N_{\mathrm{par}}} H_{ij} y_i y_j
\end{equation}
where $\{y\}$ are the deviations of the parameters $\{\xi\}$ from their best fit values,
and the derivatives in Eq.~(\ref{eq:hij}) are taken at the minimum.
An improved iterative algorithm has been devised in \cite{ref:hessian}
to evaluate the derivatives in (\ref{eq:hij}) reliably in case
of very disparate uncertainties in different directions of the multi-dimensional space
with $N_{\mathrm{par}}$ parameters describing PDFs or FFs in global QCD analyses.
We adopt this improved Hessian method also in our studies.

It is convenient to express the Hessian $H_{ij}$ in terms of its $N_{\mathrm{par}}$
eigenvectors and replace the displacements $\{ y\}$ in Eqs.~(\ref{eq:hij}) and 
(\ref{eq:chi2hessian}) by a new set of parameters $\{z\}$ related to the
eigenvector directions. If properly scaled by the corresponding eigenvalues,
surfaces of constant $\chi^2$ turn into hyper-spheres in $\{z\}$ space, and 
the distance from the minimum is given by
\begin{equation}
\label{eq:deltachi2z}
\Delta \chi^2=  \sum_i^{N_{\mathrm{par}}} z_i^2\,.
\end{equation}
Within this eigenvector representation $\{z\}$, one can straightforwardly construct 2$N_{\mathrm{par}}$ 
eigenvector basis sets of nPDFs which greatly facilitate the propagation of nPDF
uncertainties to arbitrary observables ${\cal{O}}$~\cite{ref:hessian}.
These basis sets $\{S^{\pm}\}$ correspond to positive and negative displacements 
along each of the eigenvector directions by the amount $T=\sqrt{\Delta \chi^2}$
still tolerated for an acceptable global fit.
To estimate the error $\Delta{\cal{O}}$ on a quantity ${\cal{O}}$ away from its 
best fit estimate ${\cal{O}}(\{\xi_0\})$ it is only necessary to evaluate 
${\cal{O}}$ for each of the 2$N_{\mathrm{par}}$ sets $\{S^{\pm}\}$~\cite{ref:hessian}, i.e.,
\begin{equation}
\label{eq:obserror-hessian}
\Delta{\cal{O}} = \frac{1}{2} \left[ \sum_{i=1}^{N_{\mathrm{par}}}
[{\cal{O}}(S^+_i) - {\cal{O}}(S^-_i)]^2 \right]^{1/2}\;.
\end{equation}
The simplicity of this procedure is the main advantage of the Hessian approach when compared
to the more robust, but computationally more involved and less user-friendly method
based on Lagrange multipliers \cite{ref:lagrange}.
One must keep in mind though that the propagation of PDF uncertainties in
the Hessian method has been derived under the {\em assumption} that a 
first order, linear approximation is adequate. Of course, due to the
complicated nature of a global fit, deviations, also from the
simple quadratic behavior in Eq.~(\ref{eq:chi2hessian}),
are inevitable, and error estimates based on the Hessian method 
are not necessarily always accurate.

To initiate the discussions of the nuclear modification factors $R_i^A$
and their uncertainties obtained from the global fit, we display in 
Fig.~\ref{fig:rpdf-at1} our results
for gold at the initial scale of $Q_0^2=1\,\mathrm{GeV}^2$, where we assume
a common $R_{\bar{u}}^A=R_{\bar{d}}^A=R_{\bar{s}}^A$ for all sea quark flavors
and $R_{u_{v}}^A=R_{d_{v}}^A$; see Eqs.~(\ref{eq:rval}) - (\ref{eq:rglue}),
(\ref{eq:adep}), and Tab.~{\ref{tab:para}.
Corresponding results evolved to a higher scale of $Q^2=10\,\mathrm{GeV}^2$
are shown in Fig.~\ref{fig:rpdf-at10}, where we also compare to our
previous fit \cite{ref:nds} and the recent analysis of EPS \cite{ref:eps09}.
To illustrate the $A$ dependence, nuclear modification factors in
Fig.~\ref{fig:rpdf-at10} are given not
only for gold ($A=197$) but also for beryllium ($A=9$), iron ($A=56$), and 
lead ($A=208$).
%
%
\begin{figure}[th!]
\begin{center}
\vspace*{-0.6cm}
\epsfig{figure=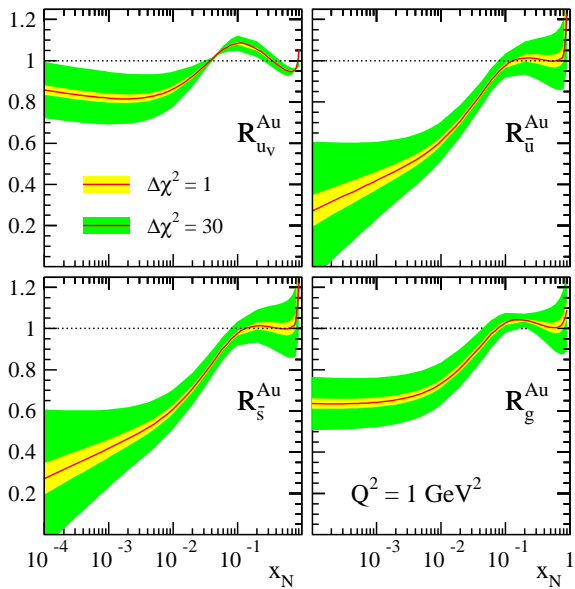,width=0.499\textwidth}
\end{center}
\vspace*{-0.5cm}
\caption{\label{fig:rpdf-at1}
The obtained NLO nuclear modification factors $R_i^{Au}(x_N)$, 
defined in Eqs.~(\ref{eq:rval}) - (\ref{eq:rglue}), 
for gold at our initial scale of $Q_0^2=1\,\mathrm{GeV}^2$.
The inner and outer shaded bands correspond to uncertainty estimates 
based on (\ref{eq:obserror-hessian}) for $\Delta \chi^2 =1 $ and
30, respectively.}
\end{figure}

Experimental uncertainties are propagated to the obtained $R_i^A$ using
the Hessian method outlined above. 
Excursions of the individual eigenvector directions resulting in a $\Delta\chi^2$ 
of 1 and 30 units are tolerated and shown as the inner and outer shaded bands,
respectively, in Figs.~\ref{fig:rpdf-at1} and \ref{fig:rpdf-at10}.
In general, $\Delta \chi^2=1$ will seriously underestimate the 
uncertainties of PDFs and FFs, and, hence, a much larger $\Delta \chi^2$ is
usually tolerated for an acceptable fit \cite{ref:dssv,Martin:2009iq,ref:dsspion,Guzzi:2011sv}.
Sophisticated dynamical criteria, see, e.g., Ref.~\cite{Martin:2009iq},
have been devised within the Hessian method to estimate a suitable $\Delta \chi^2$ corresponding 
to a, say, $68\%$ (one sigma) confidence level for a given fit, but details vary.
Reasons for deviating from the default $\Delta \chi^2=1$ are manifold and can be mainly related to
uncertainties inherent to the theoretical framework used to describe the data, which are
notoriously difficult to quantify. Examples are the choice of the factorization scale, the functional
form used to parametrize the PDFs, or unavoidable approximations curtailing the available parameter space.
Given the still rather limited amount and kinematic coverage of data taken on nuclear targets and
their relatively large uncertainties compared to data constraining free proton PDFs, we take
$\Delta \chi^2=30$, corresponding to an increase in $\chi^2$ of about $2\%$, 
for an estimate of nPDF uncertainties.
In any case, it should be kept in mind that uncertainty bands are only meaningful for combinations
of nPDFs and in kinematic regions which are actually constrained by data. Therefore, for nPDFs,
uncertainty estimates below $x_N\simeq 0.01$ should be taken with a grain of salt and merely reflect
extrapolations of the chosen functional form. In our fit this is most apparent for $R_g^A$ at small $x_N$,
where the bands shown in Fig.~\ref{fig:rpdf-at1} suggest rather small uncertainties, but only
charge and momentum conservation in Eqs.~(\ref{eq:charge}) and (\ref{eq:mom}) provide some 
limited guidance on the behavior at small $x_N$. 
 
It is worth noticing that the shape and magnitude of the nuclear modifications $R_i^A$
for a given flavor at some arbitrary scale $Q>Q_0$ depends both on the input distributions
shown in Fig.~\ref{fig:rpdf-at1} and the chosen set of reference PDFs for free protons.
Scale evolution imprints different nuclear effects on individual quark flavors
even, as in our case, one starts with $R_{\bar{u}}^A=R_{\bar{d}}^A=R_{\bar{s}}^A$ 
and $R_{u_{v}}^A=R_{d_{v}}^A$ at the initial scale $Q_0$.
These differences can be quite sizable for the strange and non strange sea 
quarks as can be inferred from comparing $R^A_{\bar{u}}$ and $R^A_{\bar{s}}$ in
Figs.~\ref{fig:rpdf-at1} and \ref{fig:rpdf-at10}.
Even for the nPDFs valence distributions, which evolve independently of the quark singlet and the
gluon, minute differences can be noticed at $Q>Q_0$ due to the different $x$ shapes of
$f^p_{u_{v}}$ and $f^p_{d_{v}}$, resulting in $R_{u_{v}}^A\neq R_{d_{v}}^A$.

%
\begin{figure*}[th!]
\begin{center}
\vspace*{-0.3cm}
\epsfig{figure=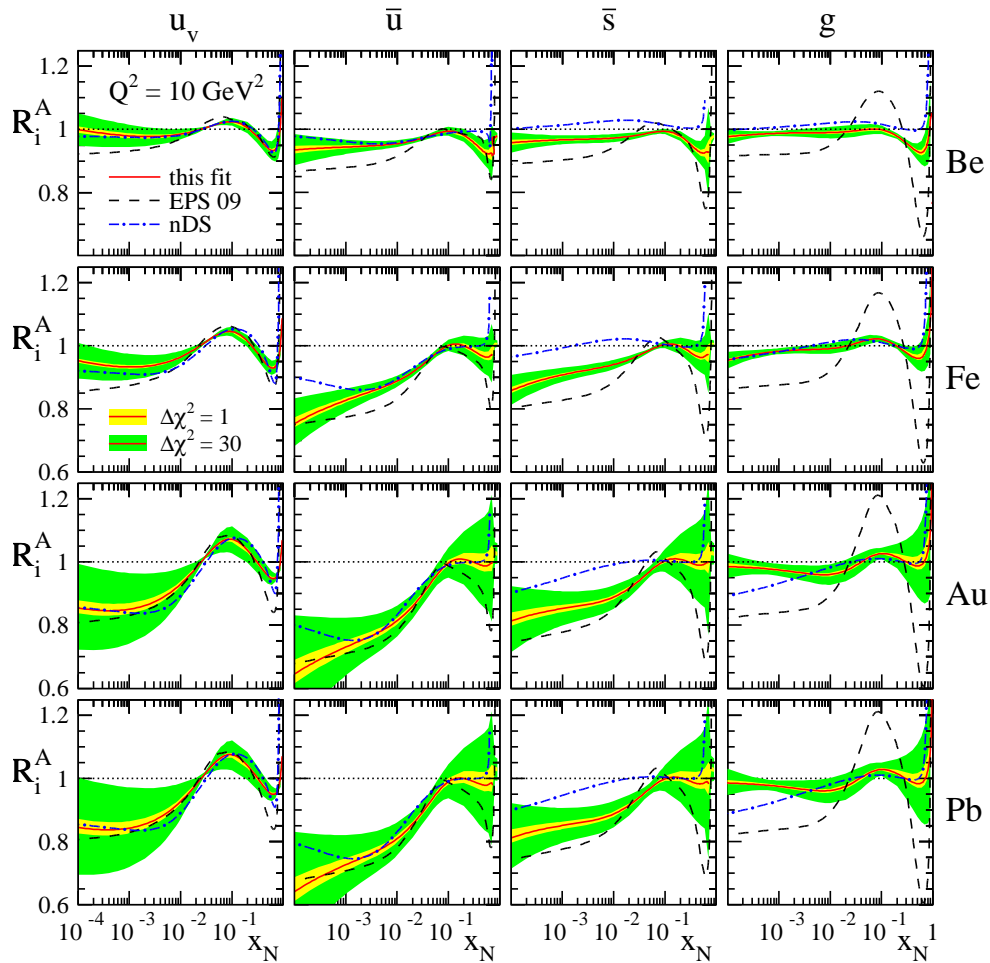,width=0.75\textwidth}
\end{center}
\vspace*{-0.5cm}
\caption{\label{fig:rpdf-at10}
Same as in Fig.~\ref{fig:rpdf-at1} but now at $Q^2=10\,\mathrm{GeV}^2$ and for four different
nuclei. Also shown are the results from our previous fit \cite{ref:nds} (dot-dashed lines)
and the recent analysis from EPS \cite{ref:eps09} (dashed lines).}
\end{figure*}
The $R^A_{u_{v}}$ in Figs.~\ref{fig:rpdf-at1} and \ref{fig:rpdf-at10} 
exhibit the expected textbook-like behavior of shadowing, anti-shadowing,
EMC effect, and Fermi motion. In the region constrained by DIS data, $x_N\gtrsim 0.01$,
uncertainties are small, even for the conservative tolerance criterion of $\Delta \chi^2=30$.
Our results also agree well with previous determinations of $R^A_{u_{v}}$ in \cite{ref:nds,ref:eps09}.
Within uncertainties, there is no evidence for any significant anti-shadowing at $x_n\simeq 0.1$ 
also for sea quarks. Again, agreement with other fits \cite{ref:nds,ref:eps09} is good, in particular, 
for $R_{\bar{u}}^A$. 
Deviations found in the nuclear modifications for strange and light sea quarks at $Q>Q_0$
are strongly influenced by the shapes of the corresponding distributions in the unbound proton 
adopted in each of the fits, which differ significantly, in particular, for the least
well determined density $f_s^p$ \cite{Martin:2009iq,Guzzi:2011sv}.
To some extent differences with previous fits \cite{ref:nds,ref:eps09} can be also
attributed to the more flexible functional form (\ref{eq:rsea})
to accommodate neutrino DIS data, as well as their impact on the fit.
Another factor is the strong correlation with $R_g^A$ at intermediate to large $x_N$.
As sea quarks are mainly constrained by DIS data at the lowest available $x_N$ and 
DY di-lepton production, uncertainties bands are smallest for $0.01\lesssim x_N \lesssim 0.1$,
and results for $x_N\lesssim 0.01$ are solely extrapolations.

As already mentioned in Sec.~\ref{sec:dau} when discussing $dAu$ data, 
our extracted nuclear modifications $R_g^A$ for the gluon density 
are expected to differ significantly from those determined by EPS \cite{ref:eps09}.
Indeed, as can be seen in Fig.~\ref{fig:rpdf-at10}, we find a much less pronounced 
anti-shadowing region around $x_N\simeq 0.1$ and EMC effect at large $x_N$ than in the EPS analysis,
mainly driven by the way in which the $dAu$ data are analyzed; see discussions in Sec.~\ref{sec:dau}
above. Differences with our previous fit \cite{ref:nds} are small, however, despite not incorporating
any $dAu$ data and defining $R_g^A$ through a convolution with free proton PDFs, see Eq.~(\ref{eq:npdfconv}).
Compared to EPS, our best fit has significantly less shadowing at $Q^2=10\,\mathrm{GeV}^2$ in
the unmeasured small $x_N$ region, but our uncertainty band clearly underestimates the true uncertainties
in this regime and is biased by the chosen functional form which is optimized to provide a
good description of the data.

%
\begin{figure}[th!]
\begin{center}
\vspace*{-0.6cm}
\epsfig{figure=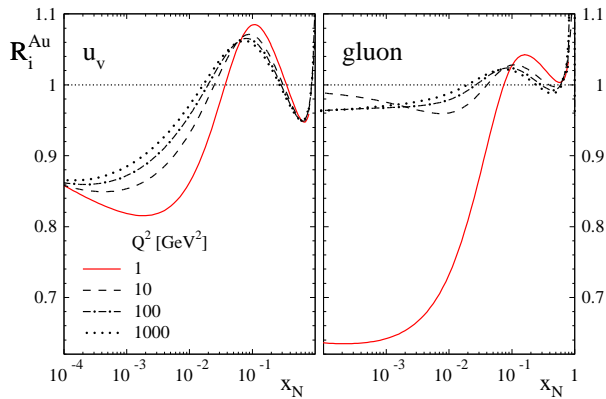,width=0.48\textwidth}
\end{center}
\vspace*{-0.5cm}
\caption{\label{fig:rv-rg-scale}
Scale dependence of the valence quark (left) and gluon (right) nuclear modification
factors for a gold nucleus as a function of $x_N$.}
\end{figure}
%
%
\begin{figure}[ht!]
\begin{center}
\vspace*{-0.6cm}
\epsfig{figure=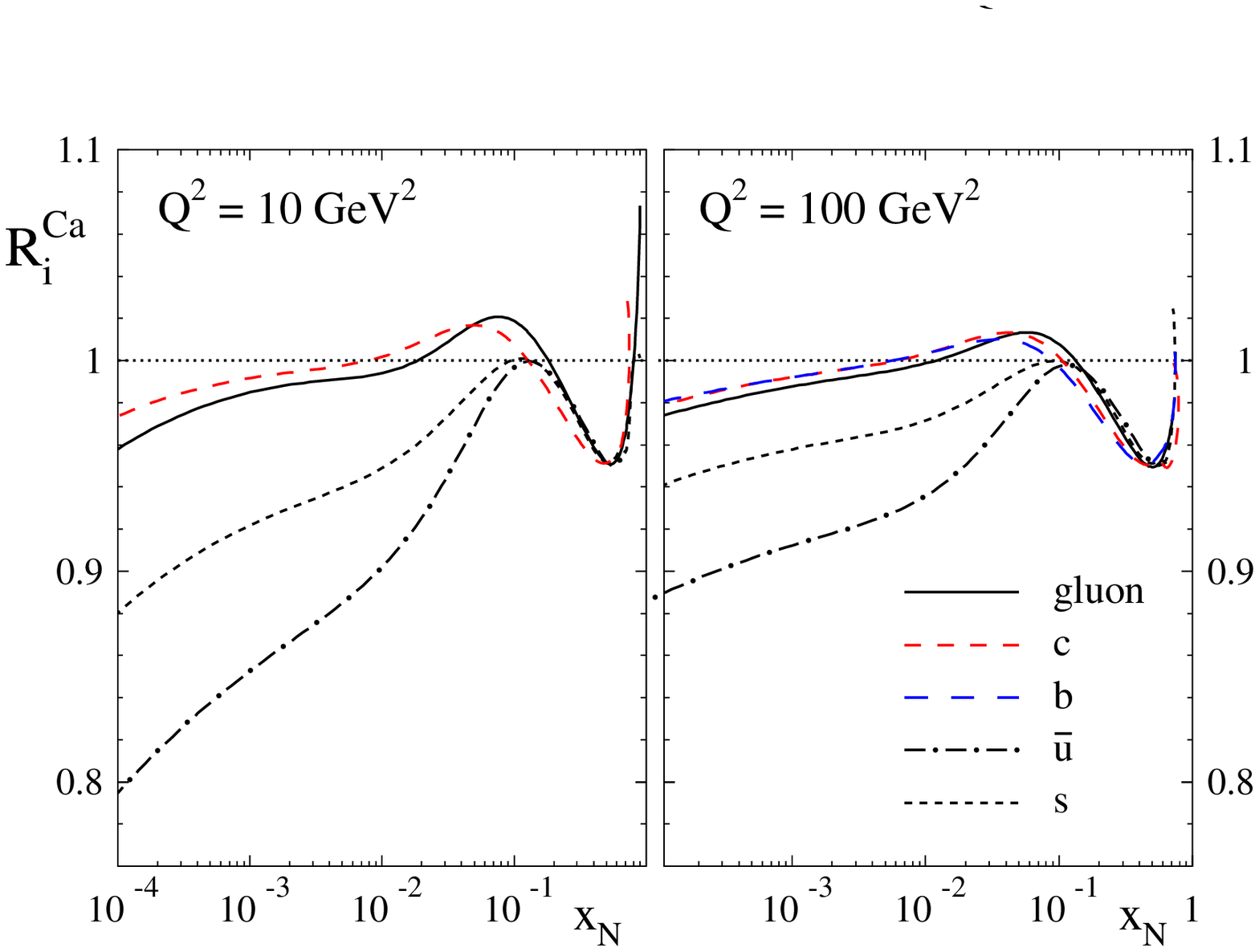,width=0.48\textwidth}
\end{center}
\vspace*{-0.5cm}
\caption{\label{fig:npdf-hq}
Typical nuclear modifications obtained for the perturbatively generated 
heavy quark parton densities compared with those of
the gluon and sea quarks for two values of $Q^2$.}
\end{figure}
It is important to notice that despite finding a comparatively moderate nuclear modification of the
gluon density at $Q^2=10\,\mathrm{GeV}^2$ in Fig.~\ref{fig:rpdf-at10}, 
shadowing is much more significant for $x_N\lesssim 0.05$ at lower scales, see, e.g.,
Fig.~\ref{fig:rpdf-at1}. 
Compared to $R_{u_{v}}^A$, $R_g^A$, and also $R_{\bar{u}}^A$,
exhibit a much more rapid scale evolution at low scales as is illustrated in Fig.~\ref{fig:rv-rg-scale}.
The large suppression of hadron yields in $dAu$ collisions at forward rapidities found
by BRAHMS and STAR \cite{Arsene:2004ux,Adams:2006uz} is essentially sensitive to
$R_g^A$ at scales around $1-2\,\mathrm{GeV}^2$ and $x_N\simeq 0.001$, 
and $R_g^A$ can be forced to describe the data without spoiling the agreement of the fit
with any other data; see also the discussion in \cite{ref:eps09}.  
As mentioned above, we refrain from doing so as the applicability of pQCD is not
guaranteed in this kinematic region and the poor knowledge of FFs and possible
medium modifications are other obstacles.
As is also apparent from a comparison of Figs.~\ref{fig:rpdf-at1} and \ref{fig:rpdf-at10},
the uncertainties on $R_g^A$ at small $x_N$ also rapidly shrink under $Q^2$ evolution, 
presumably due to gluon radiation from quarks at large $x_N$, where they are well constrained.
This was also observed in the EPS analysis, see Fig.~3 in \cite{ref:eps09}, which uses
a tolerance level of $\Delta \chi^2=50$ for 929 data points, compared to our $\Delta\chi^2=30$
for 1579 data points.

In order to illustrate also the effective nuclear modification for heavy quark flavors, we
show in Fig.~\ref{fig:npdf-hq} the ratios of the the perturbatively generated charm and 
bottom nPDFs in a calcium nucleus 
and their counterparts for a free proton from MSTW \cite{Martin:2009iq}.
Compared to the nuclear modifications of the light sea quarks and gluons, which
are also displayed in Fig.~\ref{fig:npdf-hq}, one finds that $R_c^{Ca}$ and
$R_b^{Ca}$ follow closely the $x_N$-shape of $R_g^{Ca}$. 
Such a behavior is expected as heavy quarks
are generated by gluon splitting without requiring any non-perturbative input.
The resulting $R_c^{Ca}$ and $R_b^{Ca}$ are hence fairly close to unity.
More pronounced modifications $R_g^A$ of the gluons as, for instance, obtained
in the EPS analysis \cite{ref:eps09}, would imprint larger effects also on
$R_c^A$ and $R_b^A$.
There is also an interesting hierarchy in the amount of low $x_N$ suppression, which
is stronger the lighter the quark is.

Next, we discuss an important peculiarity of our gluon nPDF, not encountered or ignored so far in
any of the previous analyses \cite{ref:nds,ref:hirai,ref:eps09,ref:schienbein}.
While LO PDFs can be assigned a physical interpretation as probabilities, at NLO and beyond,
they become scheme-dependent, non-physical quantities. In some recent fits of free proton PDFs,
including our reference set of MSTW \cite{Martin:2009iq}, the possibility of {\em negative} gluons
at small momentum fractions and scales has been entertained and is actually preferred in 
the best fit of MSTW. The $\overline{\mathrm{MS}}$ 
evolution kernels exhibit large order-by-order corrections at small
$x$ \cite{ref:pij}, resulting in rather unstable gluon distributions and huge corrections; see Fig.~57 
and the detailed discussions in Sec.~13 of Ref.~\cite{Martin:2009iq}. 
Since quarks tend to rise faster due to increasing powers of $\ln x$ in the splitting
function $P_{qg}$ at higher orders \cite{ref:pij},
gluons compensate for that by dimishing at small $x$ and $Q^2$.
%
\begin{figure}[b!]
\begin{center}
\vspace*{-0.6cm}
\epsfig{figure=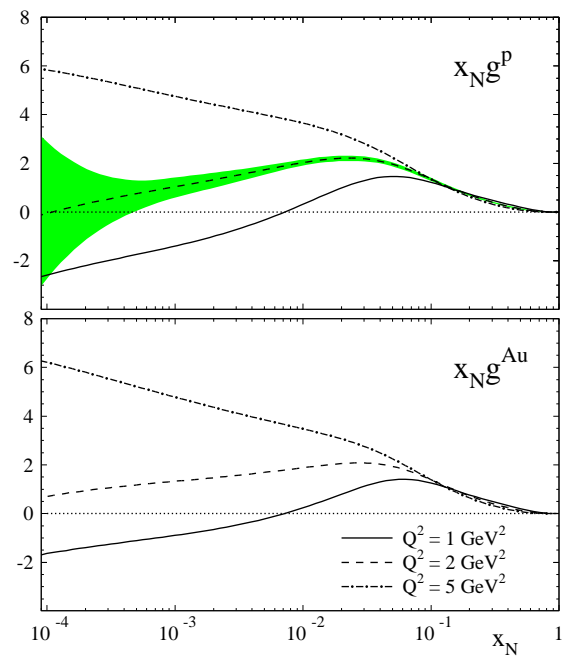,width=0.48\textwidth}
\end{center}
\vspace*{-0.5cm}
\caption{\label{fig:gluons}
Gluon distribution in a free (MSTW \cite{Martin:2009iq}) 
and bound (gold nucleus) proton at
low values of $Q^2$.
The shaded band illustrates the
$68\%$ confidence level uncertainties at $Q^2=2\,\mathrm{GeV}^2$ as estimated by MSTW.}
\end{figure}

As a result, the NLO gluon distribution in the MSTW analysis becomes valence-like at
$Q^2\simeq 2\,\mathrm{GeV}^2$ and negative at low $x$ for smaller scales as can be
inferred from Fig.~\ref{fig:gluons}. At scales $Q^2\gtrsim 2\,\mathrm{GeV}^2$, the gluon
distribution starts to exhibit the well-known strong rise at small $x$.
Negative gluons as such are not a problem as long as any physical cross section stays
positive. The DIS structure function $F_L$ is presumably the quantity which is most
sensitive to its gluon contribution. The corresponding gluonic coefficient function
also receives large order-by-order corrections at small $x$, which counter the decrease of the gluons 
and stabilize $F_L$ beyond the NLO approximation \cite{ref:flcoeff}. 
Despite having negative gluons at small $x$ and low $Q^2$, $F_L$ is well behaved
for $Q^2\gtrsim 2\,\mathrm{GeV}^2$ at NLO and down to even lower $Q^2$ at NNLO;
see Fig.~58 in \cite{Martin:2009iq}.

Since our nPDFs are tied to the free proton PDFs of MSTW through the ansatz
(\ref{eq:rval})-(\ref{eq:rglue}), our NLO nuclear gluon density inevitably 
also turns negative at small $x_N$ and $Q^2\lesssim 2\,\mathrm{GeV}^2$ as can be seen 
from the lower panel of Fig.~\ref{fig:gluons}. Since $R_g^A<1$ for small $x_N$ at scale $Q_0$, nPDFs are slightly
less negative, and we have checked that $F_L$ is positive for $Q^2\gtrsim 2\,\mathrm{GeV}^2$.
In any case, it should be kept in mind, that the region $x_N\lesssim 0.01$ is not constrained by
any experimental result for nPDFs.
Also, since $g^p$ and $g^A$ evolve differently with scale, the ratio 
$R_g$ is not well defined unless both gluon densities turn positive.
For $Q^2\lesssim 2\,\mathrm{GeV}^2$ and small values of $x_N$, the ratio $R_g$ can
have nodes for our fit.

In Fig.~\ref{fig:gluons} we also illustrate the typical uncertainties for the gluon PDF in a free proton
in the low $Q^2$ region as estimated by MSTW (shaded band). They turn out to be very sizable for $x$ values
below a few times $10^{-4}$. 
Corresponding uncertainties for the gluon distribution in a nucleus are even larger due to
the extra uncertainties introduced by $R_i^A$. Since we parametrize the nuclear modification
factors and estimate their uncertainties with respect to the best fit of MSTW, a self-consistent
propagation of uncertainties to the nPDFs is not strictly possible; see discussions in
Sec.~4 of \cite{ref:eps09}. To satisfy constraints from momentum and charge conservation,
a simultaneous fit of free and bound proton PDFs would be necessary which does not appear
to be feasible at this point given the limited experimental information with nuclear targets.

%
\begin{figure}[th!]
\begin{center}
\vspace*{0.3cm}
\epsfig{figure=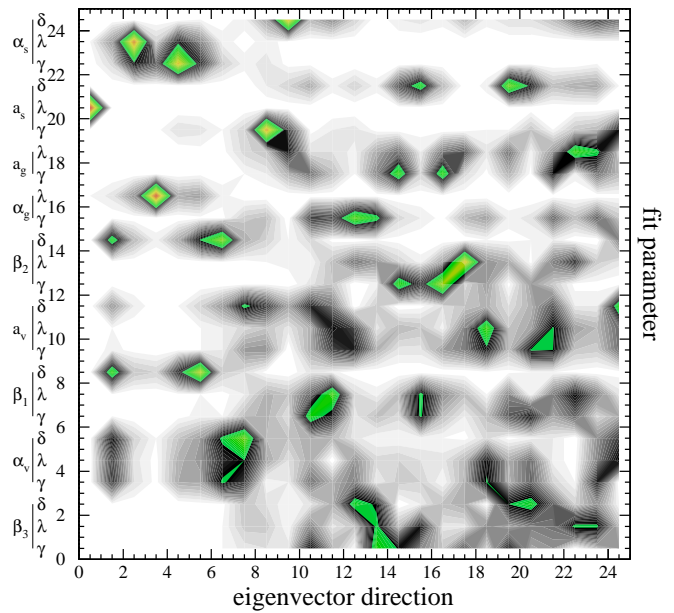,width=0.48\textwidth}
\end{center}
\vspace*{-0.3cm}
\caption{\label{fig:correlations}
Correlations between the fit parameters listed in Tab.~\ref{tab:para}
and the eigenvector directions of the Hessian matrix (see text).}
\end{figure}
Finally, for completeness, we have a closer look at the actual behavior of the $\chi^2$ function (\ref{eq:chi2})
near its minimum.
As described above, it is advantageous to work with the eigenvector directions $\{z\}$
of the Hessian matrix, where surfaces of constant $\chi^2$ are turned into hyper-spheres.
The contours in Fig.~\ref{fig:correlations} illustrate the overlap of each of the original $N_{\mathrm{par}}=25$
fit parameters $\{\xi\}$ listed in Tab.~\ref{tab:para} with the set of eigenvectors $\{z\}$.
Eigenvector 1 corresponds to the largest eigenvalue of the Hessian
matrix, i.e., the direction where $\chi^2$ changes most rapidly, and number 25 relates
to the smallest eigenvalue.
One can see that several eigenvectors have fairly strong correlations with only one or
a small group of fit parameters, while others, in particular, those corresponding to smaller eigenvalues, 
overlap with more fit parameters.
Overall, the result is not quite the ideal case, with a one-to-one correspondence
between $\{\xi\}$ and $\{z\}$, thus making it difficult to draw conclusions.
It basically reflects the complicated nature of a global analysis with minimizations
in a highly correlated, multi-dimensional parameter space and most likely the still insufficient amount of
experimental information to clearly pin down differences between sea and valence quarks
on the one hand and sea quarks and gluons on the other.
The need to parametrize not only the $x_N$ shape of the nPDFs but also their $A$
dependence further complicates the task.

%
\begin{figure}[th!]
\begin{center}
\vspace*{-0.3cm}
\epsfig{figure=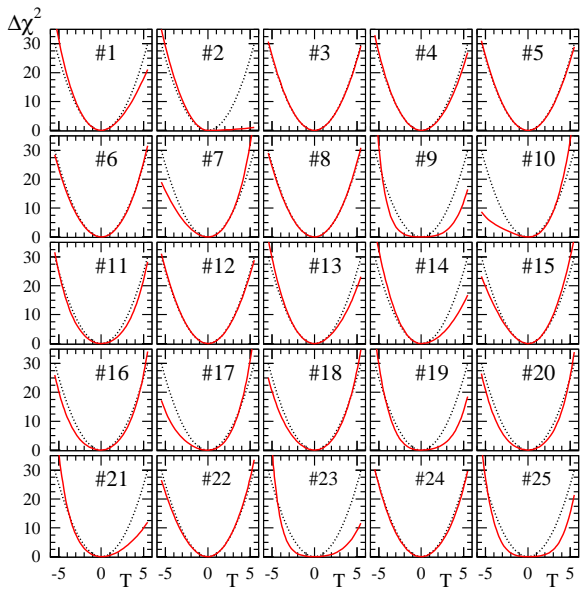,width=0.49\textwidth}
\end{center}
\vspace*{-0.5cm}
\caption{\label{fig:profiles}
Deviations (solid lines) from the expected parabolic behavior (dotted lines)
$\Delta \chi^2=T^2$ for each of the $N_{\mathrm{par}}=25$
eigenvector directions $\{z\}$ of the Hessian matrix.} 
\end{figure}
In Fig.~\ref{fig:profiles} we investigate the $\chi^2$ profiles for each eigenvector
direction. We vary one of the parameters $\{z\}$ at a time, keeping all other fixed. Of course, since
each eigenvector overlaps, in principle, with all fit parameters, as is illustrated in Fig.~\ref{fig:correlations},
the latter are all allowed to change in this procedure. The variation is done in such a way that a given 
increase $\Delta \chi^2=T^2$ is produced, and we compare the actual behavior (solid lines) for each eigenvector
direction with the parabolic
one (dotted lines) {\em assumed} in the Hessian approach. One can see from Fig.~\ref{fig:correlations}
that the quadratic approximation works reasonably well for most of the eigenvectors, with only few exceptions,
most noticeable the profile for the direction corresponding to the second largest eigenvalue. Here,
$T>0$ leads to rather tiny changes in $\chi^2$. From Fig.~\ref{fig:correlations}
one can infer that this eigenvector is mainly correlated with the parameters $\alpha_v$, $a_v$, and
$\alpha_g$. The lack of data at small $x_N$ might explain to some extent the distorted $\chi^2$ profile.
Overall, we conclude from this exercise that our eigenvector sets $\{S^{\pm}\}$ for $\Delta \chi^2=30$ produce 
reasonable uncertainty estimates in the kinematic region constrained by data, i.e., for $x_N>0.01$, 
with some caveats concerning the flavor and the quark versus gluon separation of nuclear modifications.

%
\begin{figure*}[th!]
\begin{center}
\vspace*{-0.3cm}
\epsfig{figure=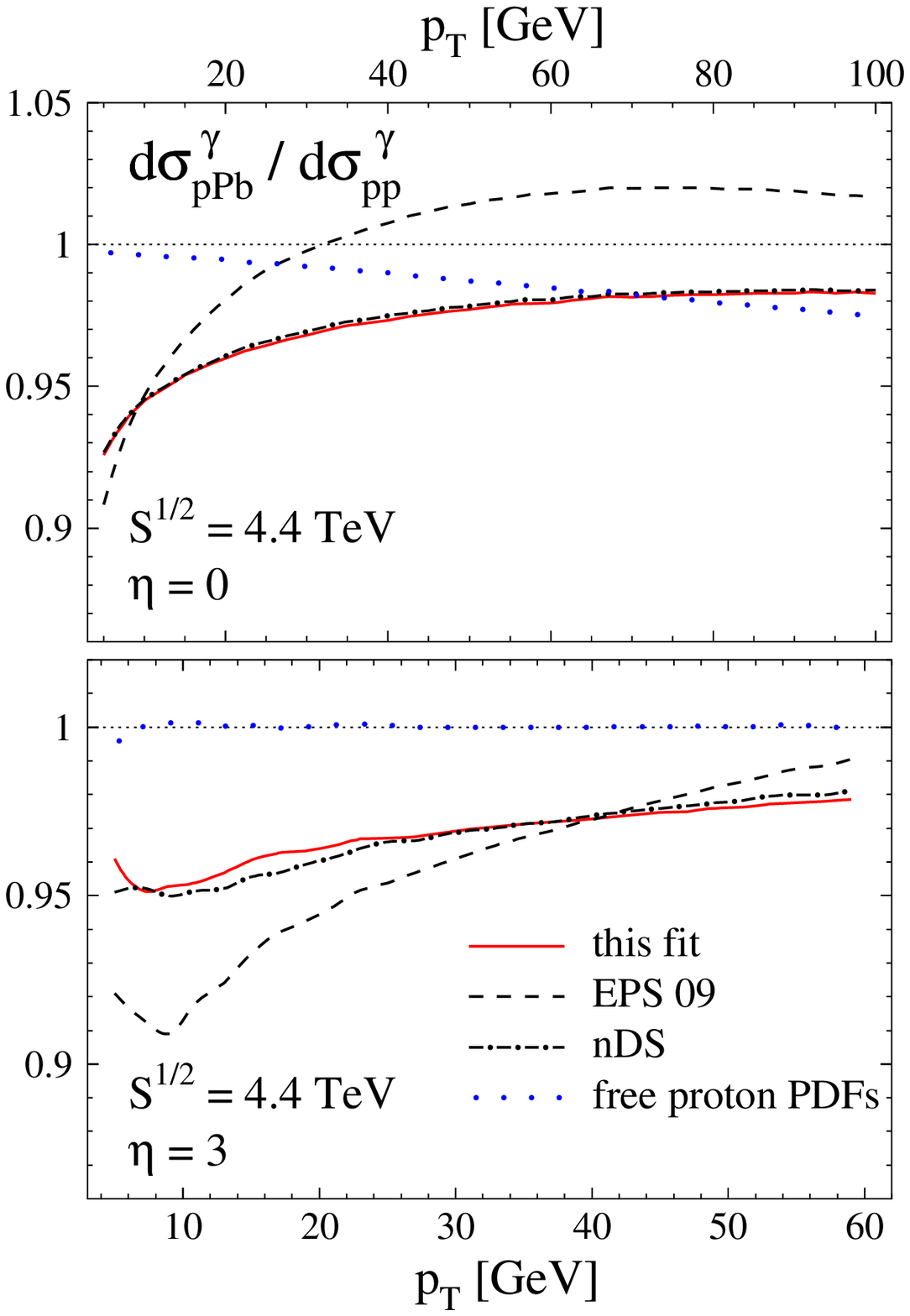,width=0.47\textwidth}
\epsfig{figure=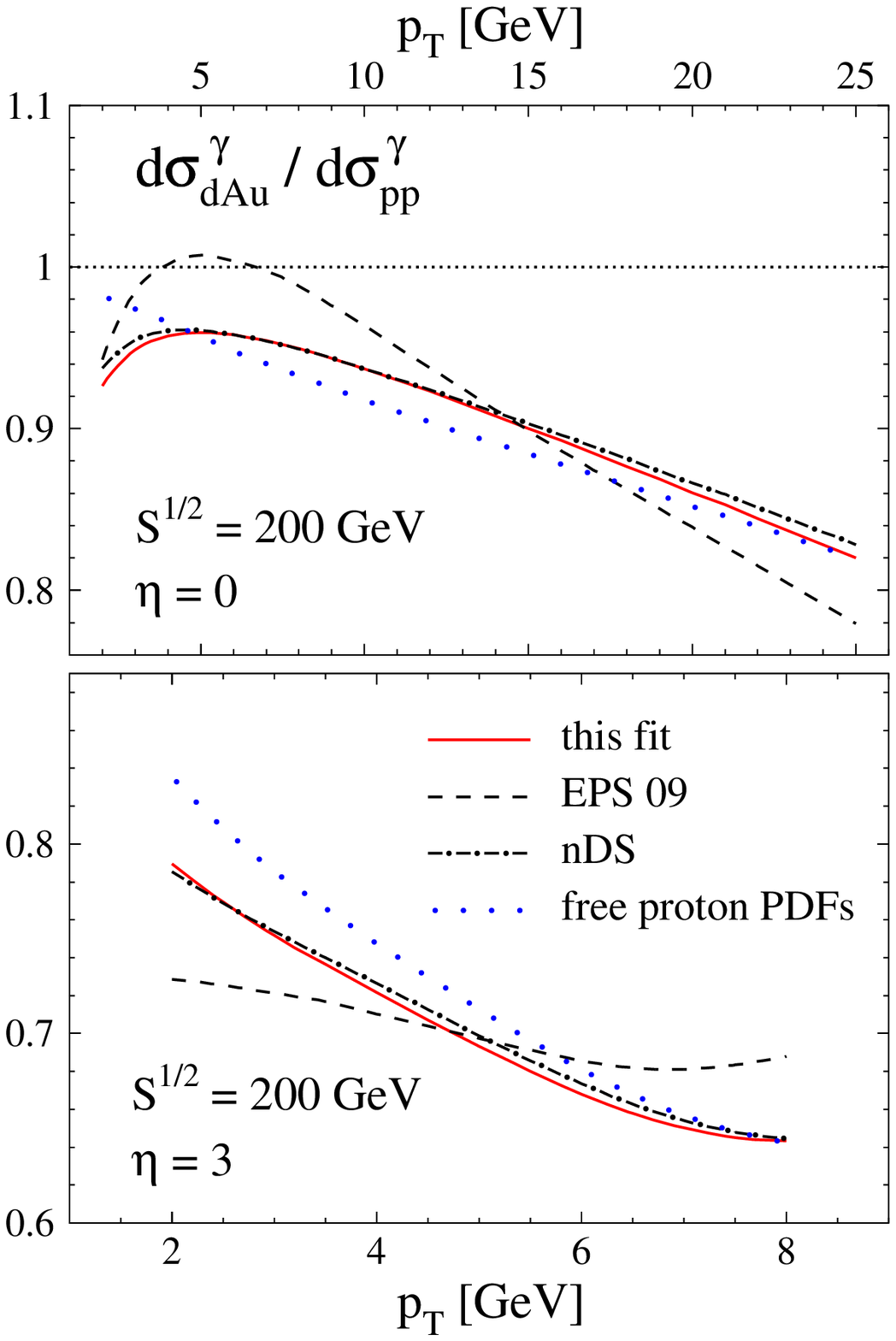,width=0.47\textwidth}
\end{center}
\vspace*{-0.5cm}
\caption{\label{fig:photons} Expectations for prompt photon production in $pPb$ (left) and
$dAu$ (right) collisions at the LHC and RHIC, respectively, for central (upper part) and
forward (lower part) photon rapidities $\eta$ using our set of nPDFs (solid lines).
Also shown are the results obtained with the nDS \cite{ref:nds} (dot-dashed lines) 
and EPS \cite{ref:eps09} (dashed lines) sets of nPDFs.
The dotted lines indicate the relevance of the isospin effect (see text).
In each case the results are normalized to the corresponding cross section in $pp$ collisions
calculated with the PDFs from MSTW.}
\end{figure*}
%
\section{Future probes\label{sec:future}}
%
As has become clear from the discussions in the previous Section, more data are
needed to further our knowledge of nPDFs to a point where one can address questions
about possible deviations from linear scale evolution or a breakdown of factorization.
The biggest obstacle for all global analyses of nPDFs is the lack of any DIS collider data
with heavy ion beams.
Measurements of the structure functions $F_2$ and, in particular $F_L$ (see
\cite{Armesto:2010tg} for a recent study), 
as well as their scaling violations for various nuclei $A$ would constrain 
the initial conditions for nPDFs in a vastly extended  range of $x_N$, similar to the one 
where the partonic structure of free protons is probed at present.
This would decisively determine also the $A$ dependence of nPDFs and, most importantly, 
challenge the currently used theoretical framework in a kinematic range where large
deviations are expected \cite{ref:nonlinear}. 
Different efforts are currently underway towards a realization of an electron-ion collider,
see Refs.~\cite{ref:eic,ref:lhec} for a status of the EIC and LHeC projects, but even
in the most optimistic scenario it will take at least another decade before first
data will emerge. 

In the meantime, interesting alternative probes are the 
rapidity dependence of inclusive prompt photon and DY lepton pair production
in $dAu$ and $pPb$ collisions at RHIC and the LHC.
In particular, yields at forward rapidities, where a large $x$ valence quark in the deuteron
(proton) interacts with a wee, small $x_N$ parton in the nucleus, may reveal novel
aspects of nPDFs. 
Despite having smaller cross sections and being experimentally more challenging,
these electromagnetic probes have the advantage of not exhibiting any sensitivity to
nuclear modifications in the final state. As we have discussed in Sec.~\ref{sec:dau},
the way how the hadronization process and possible medium modifications are modeled 
can have an impact on the obtained nuclear gluon distribution. 
Prompt photon and DY di-lepton production will hence shed light on the 
consistency of presently determined nuclear effects.

As was also mentioned above, and will be seen in our results below, 
the only theoretical complication in analyzing electromagnetic probes
is the presence of potentially significant isospin effects due to the direct coupling of photons
to the electric charge of the quarks. This makes one sensitive to, for instance, the
smaller density of $u$ quarks in a nucleus than in a free proton due to the dilution from neutrons,
which was discussed in some detail in case of prompt photon production in Ref.~\cite{Arleo:2011gc}. 
Such effects need to be taken into account when quantifying genuine nuclear modifications
for bound protons.

%
\begin{figure}[th!]
\begin{center}
\vspace*{-0.3cm}
\epsfig{figure=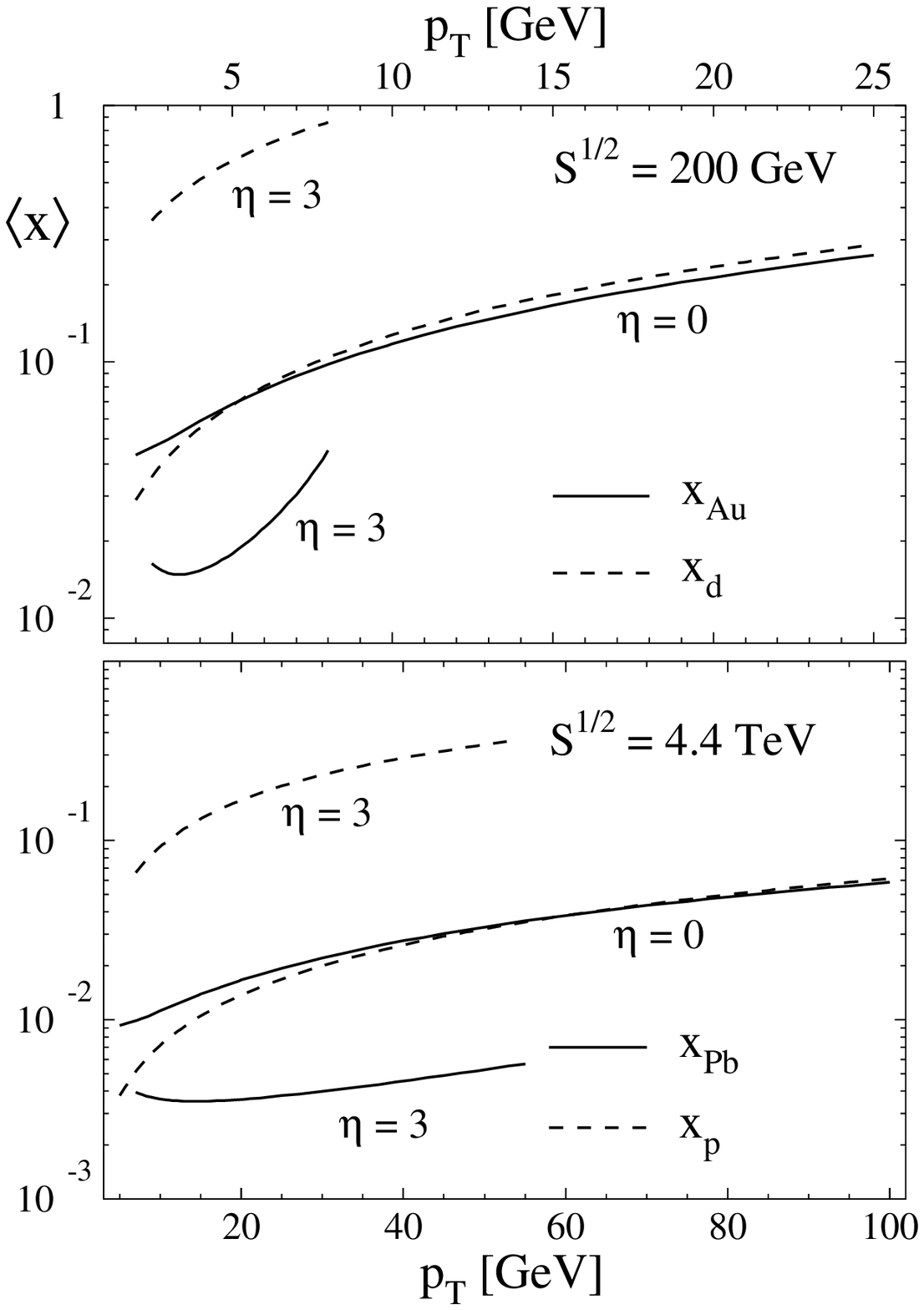,width=0.48\textwidth}
\end{center}
\vspace*{-0.5cm}
\caption{\label{fig:x-photon} 
Estimates of the average momentum fractions $\langle x_{p,d}\rangle$ and $\langle x_{Pn,Au}\rangle$ probed in
the proton (deuteron) and the lead (gold) nucleus for the cross sections shown in
Fig.~\ref{fig:photons}.}
\end{figure}
Prompt photon production has been already advocated as a probe of
the nuclear gluon density at small $x_N$ in Refs.~\cite{Arleo:2007js,Arleo:2011gc}.
Figure~\ref{fig:photons} shows expectations for prompt photon yields in $dAu$ and $pPb$ collisions,
for central and forward ($\eta=3$) photon rapidities, using our set of nPDFs (solid lines) and
normalized to the corresponding cross section in $pp$ collisions. For comparison,
the ratios are also computed with the nDS \cite{ref:nds} and EPS \cite{ref:eps09} sets of nPDFs.
All calculations are performed at NLO accuracy \cite{ref:photon}.
To extract the genuine nuclear modifications, the computed ratios should be not compared
with unity but with the dotted lines which indicate the relevance of the isospin effect.
The latter curves are obtained with free proton PDFs throughout, 
ignoring any nuclear modifications for bound protons,
and their deviation from unity is solely due to the dilution of the $u$ quark density in a 
neutron-rich nucleus where $(A-Z)>Z$.
To facilitate the discussions, Fig.~\ref{fig:x-photon} gives estimates of the 
average momentum fractions $\langle x_{p,d}\rangle $ and $\langle x_{Pn,Au}\rangle $ probed in
the proton (deuteron) and the lead (gold) nucleus 
for the results shown in Fig.~\ref{fig:photons}. The results for 
$\langle x_{p,d}\rangle $ and $\langle x_{Pn,Au}\rangle $ are obtained in the same way
as we have outlined in Sec.~\ref{sec:dau}.

At RHIC energies, central rapidity $\eta=0$, and for $p_T\simeq 10\,\mathrm{GeV}$,
one basically scans the gluon nPDF around the anti-shadowing region $x_N \sim 0.1$ and
up into the EMC effect for larger $p_T$. At $\eta=3$, one becomes sensitive to smaller
momentum fractions but not below $x_N\simeq 0.01$ already covered by other data.
Nevertheless, the pronounced differences in $R_g^A$ between the EPS and our fit,
as illustrated in Fig.~\ref{fig:rpdf-at10}, lead to characteristic differences
for $d\sigma_{dAu}^\gamma/d\sigma_{pp}^{\gamma}$, and a measurement at RHIC
will certainly help to further constrain $R_g^A$.
At the LHC, momentum fractions $x_N$ down to a few times $10^{-3}$ can be accessed 
with prompt photons produced at forward rapidities. For $\eta=0$ and $p_T\gtrsim 20\,\mathrm{GeV}$
one mainly probes the anti-shadowing peak. Again, any differences between the results
obtained with the EPS and our set of nPDFs in Fig.~\ref{fig:photons} are readily explained by the corresponding
behavior of $R_g^A$ shown in Fig.~\ref{fig:rpdf-at10}. 

As noticed from our analysis of nPDFs, Drell-Yan production provides an unique tool to disentangle 
the nuclear effects from valence and sea quark densities. At the lowest order in perturbation theory,
and keeping only the leading $u$ and $d$ quark contributions for the sake of simplicity,  
the (nuclear) cross section (\ref{eq:dyxsec}) is given by the following combination
\begin{eqnarray}
\label{eq:dy}
d\sigma_{DY}^{pA} &\propto& e_u^2 \left[ u(x_1) \bar{u}^A(x_2) + \bar{u}(x_1) u^A(x_2) \right]\nonumber \\
&+& e_d^2 \left[ d(x_1) \bar{d}^A(x_2) + \bar{d}(x_1) d^A(x_2)  \right] \, .
\end{eqnarray}
Parton distributions are probed at values of $x_{1,2}$ which depend on the invariant mass $M$ and the 
rapidity $y$ of the dilepton pair (or, equivalently, the intermediate gauge boson). 
Again, at the lowest order the momentum fractions are given by
\begin{equation}
\label{eq:xdy}
x_{1,2}=\frac{M}{\sqrt{S}}\, e^{\pm y} \, .
\end{equation}
It follows that at large {\it positive} $y$, where $x_1 \sim 1$ and $x_2\ll 1$, the cross section (\ref{eq:dy})
will be dominated by the valence distribution of the proton probed at $x_1$ 
and the sea quark $\bar{u}^A$ and $\bar{d}^A$ nuclear modified  distributions at rather low values of $x_2$. 
The measurement of the cross section ratio $d\sigma^{DY}_{pA}/d\sigma^{DY}_{pp}$ provides 
therefore direct access to the nuclear ratios $R_{\bar{u},\bar{d}}^A\,(x_2)$.
On the other hand, at large {\it negative} rapidities, distributions are probed at $x_1\ll 1$ and $x_2 \sim 1$,
and Eq.~(\ref{eq:dy}) becomes sensitive to the nuclear ratios for the valence distributions instead. 

%
\begin{figure*}[th!]
\begin{center}
\vspace*{-0.3cm}
\epsfig{figure=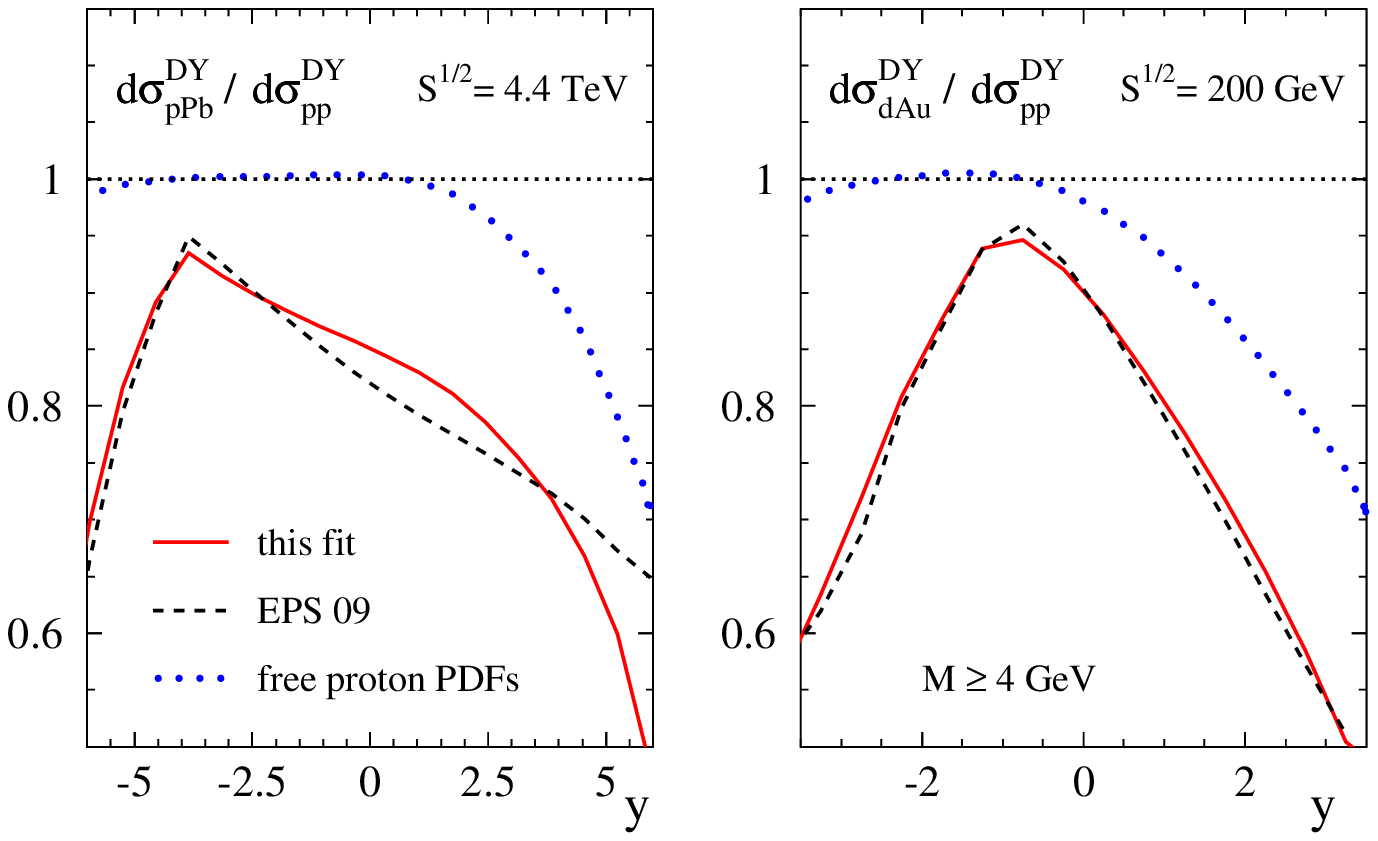,width=0.75\textwidth}
\end{center}
\vspace*{-0.5cm}
\caption{\label{fig:dy} 
Rapidity dependence for DY cross section ratios in $pPb$ (left) and $dAu$ (right) collisions at the LHC and RHIC, respectively, corresponding to invariant masses $M\geq 4$ GeV. The lines represent the expectations
for our set of nPDFs (solid), EPS \cite{ref:eps09} (dashed), and the result without considering 
nuclear effects (dotted).}
\end{figure*}
Figure~\ref{fig:dy} shows the expectations for di-lepton pair production in $pPb$ and $dAu$ collisions at the 
LHC and RHIC, respectively, for invariant masses $M\geq 4$ GeV and normalized to the corresponding $pp$
cross sections.
The calculations are performed at NLO accuracy using the code from Ref.~\cite{deFlorian:2010aa} and setting the renormalization and factorization scales as $\mu_F=\mu_R=M$. 
As in the prompt photon analysis, we include the prediction for the ratio using free proton PDFs.  
Isospin effects turn out to be rather small for negative rapidities but start to compete 
with the genuine nuclear modification at $y>0\, (2)$ for RHIC (LHC) energies. 
It is worth noticing that for RHIC kinematics it is possible to cover values of $x$ as low as $10^{-3}$ at large
forward rapidity $y=3$. 
In the case of the LHC with $\sqrt{s}=4.4$ TeV and at the same rapidity for the di-lepton pair, one can explore 
values of $x\sim 5 \cdot 10^{-5}$, where even our present knowledge of the free proton distributions 
is incomplete and will be challenged. 
It is not unexpected then, that it is in this unexplored region where we observe the largest differences 
between the predictions obtained with our nPDFs and those of Ref.~\cite{ref:eps09}, 
which otherwise agree due to their similar content of nuclear modification in the quark sector, 
as was already observed in Fig.~\ref{fig:rpdf-at10}. Clearly, measurements of DY cross section ratios
at forward rapidities will further our knowledge of nPDFs.

%
\begin{figure*}[th!]
\begin{center}
\vspace*{-0.3cm}
\epsfig{figure=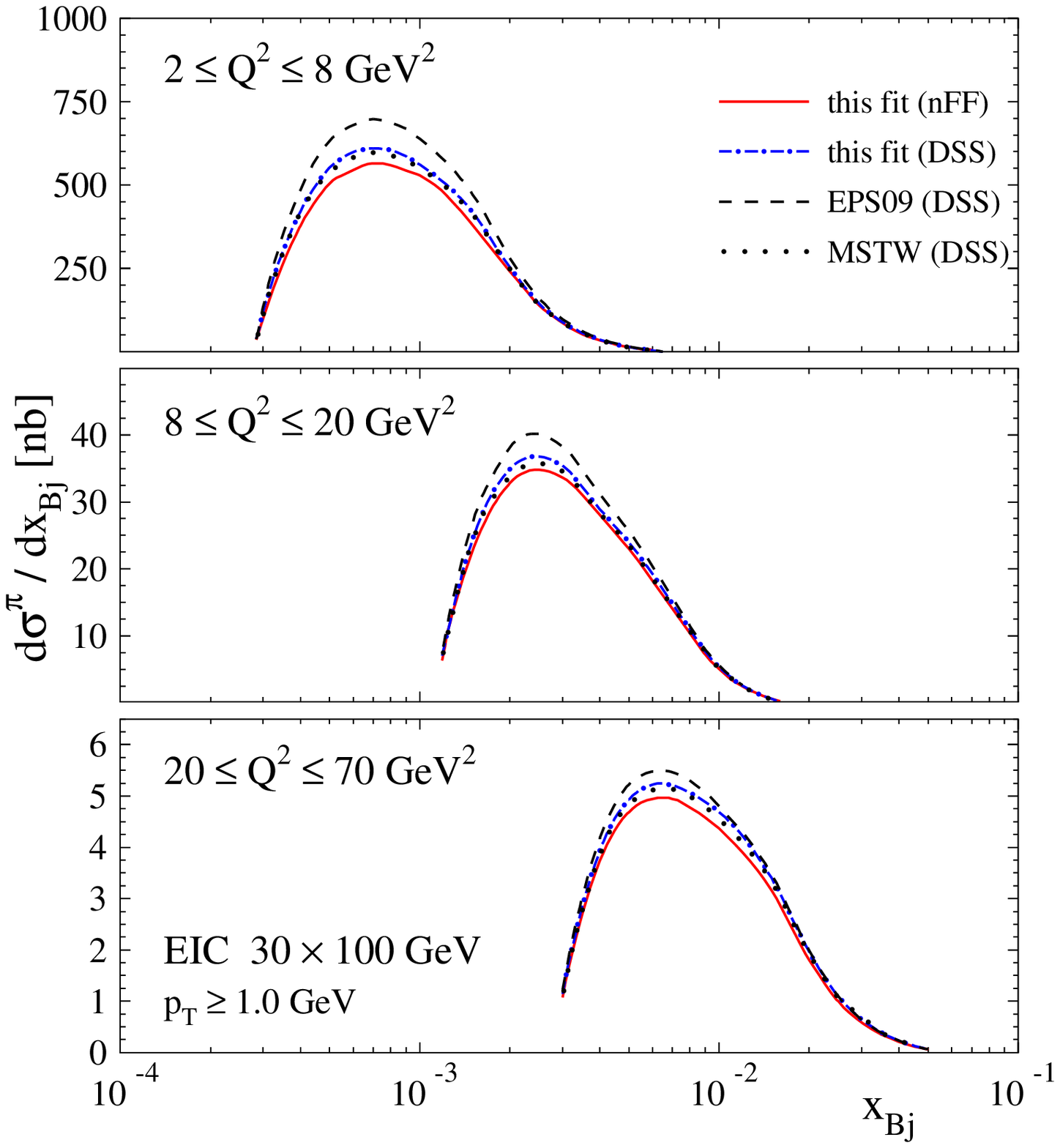,width=0.44\textwidth}
\epsfig{figure=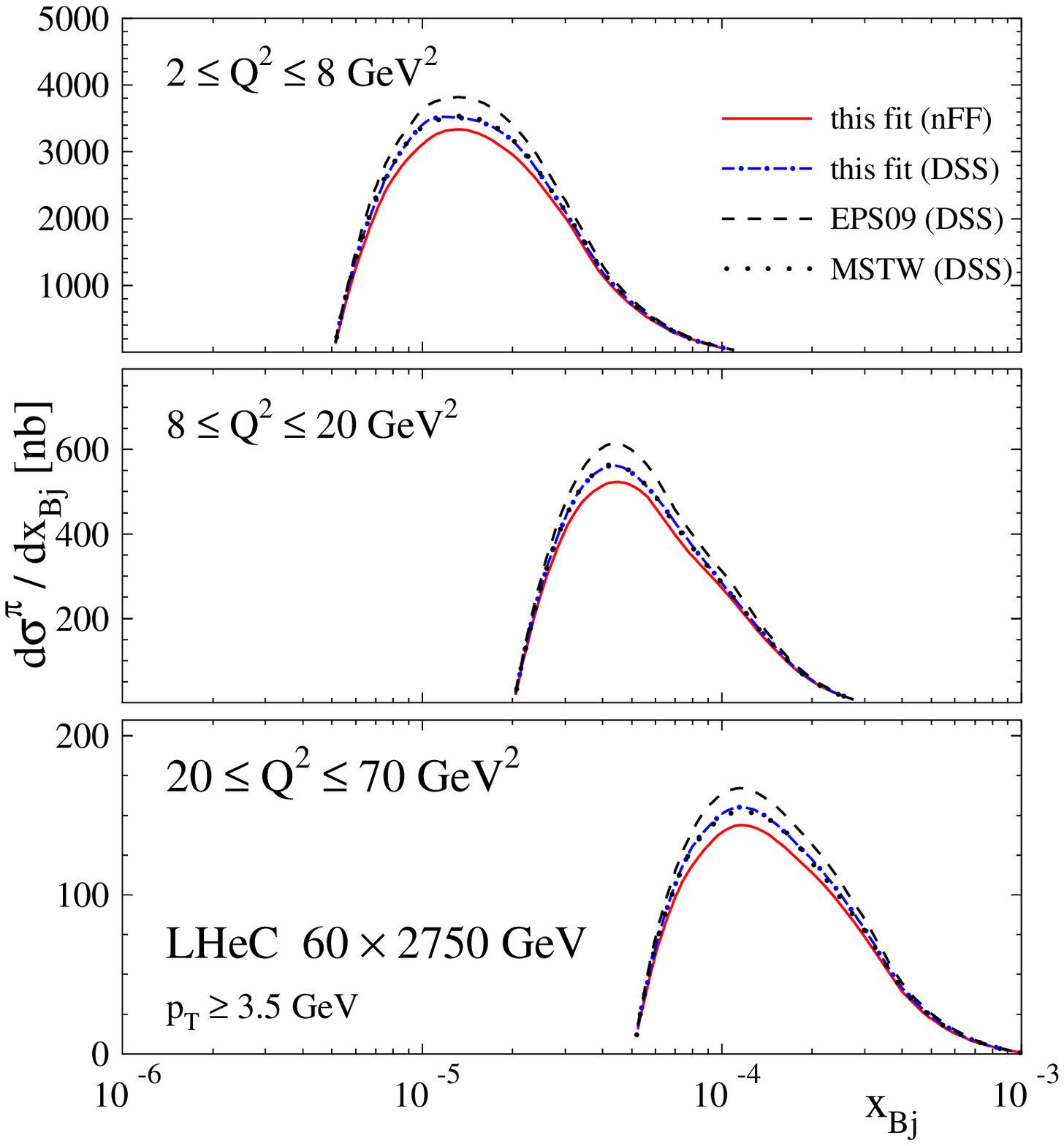,width=0.44\textwidth}
\end{center}
\vspace*{-0.5cm}
\caption{\label{fig:dishadron} Neutral pion production in DIS in different bins of $Q^2$
as a function of Bjorken $x$ for four combinations of (n)PDFs and FFs.
The solid lines represent the result obtained with our best fit of nPDFs and the medium modified
FFs of \cite{ref:nff}.
{\bf{left:}} Collisions of $30\,\mathrm{GeV}$ electrons and 
$100\,\mathrm{GeV}$ per nucleon gold ions at an EIC \cite{ref:eic}. The $p_T$ of the pion
is restricted to be larger than $1\,\mathrm{GeV}$. {\bf{right:}} Collisions of $60\,\mathrm{GeV}$ electrons and 
$2.75\,\mathrm{TeV}$ per nucleon lead ions at the LHeC \cite{ref:lhec} and $p_T>3.5\,\mathrm{GeV}$.}
\end{figure*}
Finally, we look into the production of hadrons in lepton-nucleus DIS which
constitutes an excellent benchmark for different aspects of nuclear effects
both in the initial and in the final-state.
The process is crucially sensitive to the three main ingredients of a pQCD calculation: 
the effective parton content of the nuclei, the mechanism for partons fragmenting
into the detected final-state hadron in a nuclear medium, 
and additional parton radiation before and after the interaction with the electromagnetic probe. 
The relevant pQCD framework is well known up to ${\cal{O}}(\alpha_s^2)$  
\cite{Daleo:2003jf,Daleo:2003xg}, and the phenomenological consequences of QCD
corrections have been studied in detail in \cite{Daleo:2004pn}.  
By choosing appropriate kinematical cuts, one can enhance different partonic
subprocesses which, in turn, might be affected differently in a nuclear medium.

In Ref.~\cite{Aktas:2004rb} the H1 collaboration presented a measurement for 
neutral pion production in $e^+p$ collisions at a c.m.s.\ energy of about $\sqrt{s}=300\,\mathrm{GeV}$.
The $\pi^0$'s were required to be produced within a small angle $\theta_{\pi} \in[5^o,25^o]$
from the proton beam in the laboratory frame, 
with an energy fraction $z_{\pi}=E_{\pi}/E_P>0.01$, and $2.5<p_T<15\,\mathrm{GeV}$.
The data confirmed previous measurements which suggested that pQCD predictions
at LO accuracy underestimate the cross section at low $x_B$ \cite{H1viejo} whereas
expectations based on BFKL dynamics \cite{BFKL} or on the parton content of
virtual photons \cite{Jung} yielded a better agreement.
The disagreement between the H1 data and LO estimates based on ${\cal{O}}(\alpha_s)$ 
cross sections convoluted with LO PDFs and FFs could be as large as an order of magnitude, 
depending on the particular kinematical region. 
Since this is much larger than typical higher order corrections (``$K$ factor'')
for such kind of process, it was suggested that the data indicate the onset of possible
non-linear effects in the scale evolution.
In Ref.~\cite{Daleo:2004pn} it was pointed out, however, that the particular set
of kinematic cuts implemented in the H1 analysis strongly suppresses the LO
contributions such that most of the observed cross section is indeed due to
the $\gamma^*+g\rightarrow g+q+\bar{q}$ channel, which only opens up at 
${\cal{O}}(\alpha_s^2)$, thus explaining the large $K$ factor.
More specifically, gluon initiated processes in which the pion is produced
from a fragmenting gluon were found to be dominant.

Performing similar measurements at a future electron-nucleon collider \cite{ref:eic,ref:lhec}
would allow one to probe nPDFs, possible medium modifications in the FFs, as well as
the validity of standard linear scale evolution and collinear factorization.
In Fig.~\ref{fig:dishadron} we show expectations for the neutral pion cross section
in different bins of $Q^2$, obtained with different combinations of (n)PDFs and (n)FFs
for both EIC (left panel) and LHeC (right panel) kinematics.
In case of an EIC \cite{ref:eic}, we assume collisions of a $30\,\mathrm{GeV}$ electron
beam with a $100\,\mathrm{GeV}$ (per nucleon) gold nucleus and similar cuts as in the H1 experiment
except that the transverse momentum of the pion is now allowed to go down to $1\,\mathrm{GeV}$.
For the LHeC, our results refer to collisions of $60\,\mathrm{GeV}$ electrons with
$2.75\,\mathrm{TeV}$ per nucleon lead ions.
Interestingly, for both experiments the predictions based on EPS nPDFs \cite{ref:eps09}
and ordinary vacuum FFs from DSS \cite{ref:dsspion} show a clear enhancement relative
to the expectations without any nuclear effects based on MSTW PDFs, which are
largely indistinguishable from the results obtained with our new set nPDFs and DSS FFs.
Using medium modified FFs \cite{ref:nff} instead leads to a suppression relative to the
MSTW(DSS) results.

\section{Summary and Conclusions}
%
We have performed an up-to-date determination of parton densities in nuclei \cite{ref:npdf} using an 
extended set of data for different observables involving nuclear targets  
and a modern set of parton distributions for free protons as reference. 
The resulting nPDFs are defined at NLO accuracy in QCD 
and in a general mass variable flavor number scheme.  
The determination of nPDFs includes error estimates obtained within the improved Hessian
method for $\Delta \chi^2=30$ and a collection of alternative, eigenvector sets of nPDFs that allow one 
to propagate these uncertainties, in principle, to any desired observable depending on these distributions.

Our results are fully consistent, within uncertainties, with a previous determination 
of nPDFs in Ref.~\cite{ref:nds} based on a much more limited set of data, except for the nuclear
modifications of the strange quark distribution, mainly due to significant 
changes in the underlying free proton reference density.
The nuclear modifications for gluons are still found to be rather moderate in the entire range of 
momentum fractions, despite including novel experimental results from $dAu$ collisions.
Noticeable deviations to Ref.~\cite{ref:nds} are only found towards small values of $x$ and are
due to extrapolations outside the kinematic region constrained by data.
We have also presented nuclear parton densities for charm and bottom quarks, that were ignored
in the previous analysis. They are generated radiatively, i.e., without any additional free parameters, 
from the gluon and light quark distributions in a general mass variable flavor number scheme. 

At variance with Refs.~\cite{ref:schienbein,Kovarik:2010uv}, we find no conflicting patterns of
nuclear modifications for neutral and charged current deep-inelastic scattering data.
We notice, however, that the latter set of data have an important impact on the shape of the 
extracted nuclear sea quark densities whose modification
factors appear to be significantly different to those found for the valence quarks. 

Compared to the fit in Ref.~\cite{ref:eps09}, which also includes some of the available
inclusive hadron production data in $dAu$ collisions from RHIC, we account also for possible
nuclear modifications in the hadronization process, which are known to be sizable 
in multiplicity ratios in SIDIS,
and refrain from assigning an inflated weight $\omega_{dAu}$ for this subset of data in the fit.
The resulting nuclear gluon density in the relevant $x$ region constrained by $dAu$ data 
differs considerably from the one obtained in the EPS analysis \cite{ref:eps09}. 
Compared to the latter fit, which is
characterized by an anti-shadowing and EMC effect considerable larger than those for quarks, our
gluon density exhibits only moderate nuclear corrections.
The use of standard vacuum rather than modified fragmentation functions in our global analysis 
leads to a marginally poorer quality of the fit well inside the tolerated increase in $\chi^2$.  
 
Uncertainties in the nPDFs extraction are still found to be rather large, in spite of the 
inclusion of additional data sets, in particular, when compared to the present knowledge of 
PDFs for free protons. As always, the estimated uncertainty bands depend on the chosen
tolerance criterion, for which we use $\Delta \chi^2=30$, and are only trustworthy in
the region of momentum fractions constrained by data. Extrapolations below $x_N\simeq 0.01$
depend mainly on the functional form used in the fit and do not reflect the true uncertainties. 

Finally, we have presented expectations based on the obtained set of nPDFs 
for some promising hard probes comprising prompt photon and forward DY di-lepton production
at RHIC and the LHC. These measurements are expected to further our knowledge of nuclear
modifications and test their universality. Compared to hadron or jet production at RHIC or the LHC, 
electromagnetic probes have the advantage of being independent of possible medium modifications
of the final state.
The biggest obstacle in current determinations of nPDFs is, however, the complete lack of 
collider data for DIS off nuclear targets, which would constrain nPDFs down to considerably lower
values of momentum fractions than present fixed-target data.
It is in this kinematic regime of high gluon density where one primarily expects non-linear effects in the 
scale evolution and a breakdown of standard collinear factorization.
The science case for a future electron-ion collider such as an EIC or the LHeC is currently under
review but first data are at best expected in a decade from now.

\section*{Acknowledgments}
%
We thank J.\ Bl\"{u}mlein, A.\ Hasselhuhn, S.\ Moch, and S. Alekhin
for their support concerning the heavy flavor Wilson coefficients and
W.\ Vogelsang and G.\ Watt for their help 
with the calculations of the prompt photon yields 
and the MSTW PDFs, respectively. We are also grateful to N.\ Armesto, H.\ Paukkunen and C.\ Salgado 
for useful discussions.
We thank D.\ Barmak for his participation in the initial stage of this work.
D.dF.\ acknowledges support from the Pauli Center for Theoretical Studies (Z\"urich).
M.S.\ acknowledges support by the U.S.\ Department of Energy under contract number DE-AC02-98CH10886.
This work was partially supported by CONICET, ANPCyT, UBACyT, and by the Research Executive Agency (REA) 
of the European Union under the Grant Agreement number PITN-GA-2010-264564 (LHCPhenoNet). 



\begin{thebibliography}{99}
\bibitem{ref:npdf} The best fit and the Hessian eigenvector sets of our nPDFs are available upon request from the authors.
\end{thebibliography}
\end{document}